\documentclass[12pt]{article}
\usepackage{graphicx}

\begin{document}

\title{Classical and Semiclassical Studies of the Hydrogen Molecule}

\author{A. L\'opez-Castillo \\  {\small \it Departamento de Qu\'{\i}mica,
Centro Universit\'ario FIEO (UNIFIEO)} \\
{\small \it Osasco, SP, 06020-190 Brazil}}
\date{\today}
\maketitle

\begin{abstract}

The hydrogen molecule contains the basic ingredients to understand the chemical
bond, i.e, a pair of electrons. We show a step to understand The Correspondence
Principle for chaotic system in the Chemical World.

The hydrogen molecule is studied classically as an extension of the helium atom.
Several types of orbits were found for two-fixed-centers system (hydrogen
molecule) starting from some known orbits for one-fixed center one (helium
atom), e.g., one dimension orbits as pendulum and axial and also Bohr and
Langmuir's orbits~\cite{ALC,ALC1}.

The classical stability and the single quantization of some one-dimensional
periodic orbits  are shown. These orbits are used to make a global quantization
of that molecular chaotic system. We discuss the importance of those periodic
orbits in the comprehension of the nature of chemical bond.

\end{abstract}


\newpage


\section{Introduction}

Bohr's model for the hydrogenlike atoms gives the correct (exact) spectrum for
such systems and so was an important paradigm in the development of the
so-called old quantum theory. However, the extension of that model to other
non-hydrogenlike atoms is non-trivial.

Similar ideas were soon applied to the helium atom (two-electron
system)~\cite{VV}, but the old quantum theory cannot describe the helium atom in
the same accuracy level of the hydrogen atom. Essentially, the inaccuracy is
not an exclusive manifestation of the old quantum theory, but classical theory.
A three-body system is generically a non-integrable system, it is not possible
to find the necessary number of integrals of motion as for two-body system
(integrable). A nonintegrable system can show several behavior patterns
depending on the initial conditions, i.e., a (quasi-)regular, chaotic or mixing
behaviors~\cite{MG}. The regular or integrable system obeys special conditions
which allows the application of the basic principles of the old quantum theory
rules. However, that theory has some incorrect statements, e.g., the inability
to treat the conjugated points along classical trajectories~\cite{MG}. The
correct version of the old quantum theory is denominated semiclassical theory.

Important applications of semiclassical theory can be seen in the studies of the
connection between the classical mechanic and quantum one (correspondence
principle) when the classical system is chaotic~\cite{MG,AMOA}.

The helium atom can be solved in a high degree of accuracy with quantum
mechanics with ``brute force.'' That system was studied at the end of the
seventies~\cite{LP} with a semiclassical theory obtaining good results. A decade
before, the semiclassical theory was developed by Gutzwiller with application to
the quantization of chaotic systems~\cite{MG}. Meeting those conditions, in the
beginning of the nineties, the helium atom was extensively studied with
semiclassical theory~\cite{WRT,TRR}. Several important results were managed for
helium atom in one dimension based on classical properties of the periodic
orbits.

The semiclassical theory can also be applied to molecules. Of course, the
molecular systems are remarkably more difficult to understand in comparison to
atoms. It is possible to make some extension of those atomic studies in
order to improve the knowledge in the molecular system and principally the
chemical bond.

Since Dalton's time, or even before, people try to understand the nature of the
(chemical) bond among atoms or, in other words, the molecular structure. Bohr
was the first to try to describe the chemical bond under a physical
point-of-view of the hydrogen molecule (H$_{2}$) and other systems~\cite{NB}.
Bohr made an original proposal to describe the chemical bond of the hydrogen
molecule. He proposed that the valence electrons (outer electrons) were
localized in a ring around the internuclear axis representing the chemical bond.
This was the first physical interpretation of the chemical bond. That
description was made before the {\it ad hoc} Lewis, Langmuir and Thomson's
theories~\cite{HK}.

The extension of such studies of the helium atom to molecular systems seems
natural through the study of H$_{2}$. A study of H$_{2}$ system begins as a
three-body (restricted four-body) problem where the nuclear charge in H$_{2}$ is
distributed in two points (two coupled Kepler problems of two centers). This
system of two central forces can be understood as a system between the helium
atom and two hydrogen atoms or ions.

The simplest molecule is the H$_{2}^{+}$. The molecule-ion H$_{2}^{+}$ in
adiabatic approximation is an integrable system as the isolated hydrogen
atom. Of course, the H$_{2}^{+}$ as a three-body system is a nonintegrable
problem. W.Pauli tried to calculate the energy formation of the hydrogen
molecule-ion (H$_{2}^{+}$) applying the (incorrect) old quantum theory
(Bohr-Sommerfeld quantization)~\cite{WP}. The first correct semiclassical
quantization was presented at the end of the seventies~\cite{SR}. However, the
H$_{2}$ molecule, even more so than H$_{2}^{+}$, contains the basic ingredients
for understanding the chemical bond, i.e., an electron pair.

The hydrogen molecule is the natural choice to start the study of the chemical
systems since it is the simplest problem where the chemical aspects can be
studied in full details.

The problem of the hydrogen molecule belongs in no-person's land between
Chemistry and  contemporary Physics. This molecule is very simple conceptually
considered in the point-of-view of the present Theoretical Quantum Chemistry and
very complex over first principles of the present Physic. On the other hand, we
try to show that the hydrogen molecule is a capital problem in order to
understand the fundamental matter structure and the quantum mechanics.

Semiclassical studies of~H$_{2}$ with few types of symmetric Bohr and Langmuir's
orbits  in two dimension (2-D) were considered~\cite{ALC,ALC1}. Those studies of
H$_{2}$ were made with few 2-D periodic orbit filling the 3-D. The full 3-D
problem is generically impossible to deal with. However, if one uses chemical
and physical arguments, i.e., disposition of electrons in electron-pair,
symmetries, etc, some special periodic orbits can be found and an incipient
quantization with few orbits can be achieved.

These semiclassical studies can be considered to understand the chemical bond.
It is known that the chemical bond is formed by decreasing of kinetic energy in
the chemical bond direction with accumulation of the electronic charge in the
internuclear region~\cite{KR}. We show with classical and semiclassical physics
that the perpendicular plane to the axial direction is the most important region
to describe the chemical bonding.

Here, we show a modest first step to understand ``the Correspondence Principle''
in chemical world that treats essentially nonlinear systems.

\section{Theory and Models}

H$_{2}$ is the simplest molecule in the chemical context, but it is a very
complex system  under some basic principles, i.e., H$_{2}$ is nonintegrable,
multidimensional and singular. The hydrogen molecule may be studied in the
context of nonlinear dynamics, its complex dynamics can only be studied by
nonperturbative theory.

The H$_{2}^{+}$ is the starting model to study the H$_{2}$ system as shown
below.

\subsection*{1-D two-center Kepler problem - H$_{2}^{+}$}

The Classical Hamiltonian for H$_{2}^{+}$ system in 1-D or two-centers Kepler
problem (one electron and two protons) is given by

\begin{equation}
\label{Hamiltonian1}
H_{H_{2}^{+}} = \frac{P^{2}}{2} - \frac{1}{|z-R/2|} - \frac{1}{|z+R/2|}.
\end{equation}

The Hamiltonian~(\ref{Hamiltonian1}) shows important numerical problems in the
integration  of the classical Hamilton equations since the potentials diverge
on $|z|=R/2$. However, the phase space expansion technique can solve that
numerical problem. The time $(t)$ and the (quasi-)energy $(E)$ are treated as
canonical variables in that technique. We need firstly to define a Hamiltonian
($\Gamma$) identically null. $\Gamma$ can be multiplied by the inverse of the
singular function in Hamiltonian~(\ref{Hamiltonian1}). (Such technique is the
reguralization procedure used in Celestial Mechanics~\cite{VS}.)

The regularized Hamiltonian can be achieved as shown below.
$$
\Gamma^{*} = P_{t} + H(z,P,t) = 0 ,
$$
where $P_{t}$ = -E = - $H(z,P,t)$.

The new time variable $(t^{*})$ is defined as
$$
dt/dt^{*} \equiv dq_{t}/dt^{*} = \partial \Gamma^{*}
/ \partial P_{t} =1 ,
$$
if
$$
\Gamma = f(z,P,t) \Gamma^{*} ,
$$
$t^{*'}$ variable is
$$
t^{*'} = f^{-1}(z) ( t+ cte ) ,
$$
where all divergence problem is transfered to determination of $t^{*'}$,
i.e., in the divergence region $t^{*'}$ becomes slower than $t$.

The function $f(z)$ that regularizes the Hamiltonian~(\ref{Hamiltonian1}) is

\begin{equation}
\label{regfunc}
f(z)=|{(R/2)}^{2} - z^{2}|
\end{equation}

and $\Gamma$ is

\begin{equation}
\label{gamma}
\Gamma = |{(R/2)}^{2} - z^{2}| P_{t} +  |{(R/2)}^{2} - z^{2}| P^{2}/2
- (R,2|z|),
\end{equation}
where $({R,2|z|})$ is the step function, i.e., for $|z|<R/2$ the function
$({R,2|z|})$ is $R$ with $|{(R/2)}^{2} - z^{2}|=[{(R/2)}^{2} - z^{2}]$
and for $|z|>R/2$ it is $2|z|$ with $|{(R/2)}^{2} - z^{2}|=[z^{2}-{(R/2)}^{2}]$. 
The Hamiltonian~(\ref{Hamiltonian1}) is separable in two regions (``outside'' 
and ``inside'' of the molecule) and the regularized Hamiltonian maintains
similar property.

It is necessary to make a canonical transformation in the Hamiltonian $\Gamma$
in order to obtain the Hamilton equations. Such transformation changes the
finite product $[{(R/2)}^{2}-z^{2}]P^{2}$ ($P$ diverges from $z=R/2$) to a
single canonical variable.

The ``outside'' regions are obtained with ($|z|>R/2$) and so

$$
z^{2} = {(R/2)}^{2} + \chi^{2},
$$
with $(R/2) < |z| < + \infty ,$ and $0 < |\chi| < + \infty $.

The final reguralized Hamiltonian is

\begin{equation}
\label{Hamiltonian2}
\Gamma ^{'} = \Gamma / z^{2} = 1/2 P_{\chi}^{2} + \frac{\chi^{2}P_{t}}
{[(R/2)^{2}+\chi^{2}]} -\frac{2}{[(R/2)^{2}+\chi^{2}]^{1/2}}=0,
\end{equation}
where $P_{\chi} = \chi P/ [(R/2)^{2}+\chi^{2}]^{1/2}$ and the generating
function~\cite{HG} $F_{3} = -P[{(R/2)}^{2} + \chi^{2}]^{1/2}$.

$\Gamma ^{'}$ is similar to a harmonic oscillator near the collision
electron-proton ($\chi<<R/2$). The hydrogen atom  (one-center problem) is
exactly a harmonic oscillator for similar regularization procedure~\cite{VS}.

The ``inside'' region is obtained with ($|z|<R/2$) and the $\Gamma$ is
given below with $2z/R = \cos \theta$.

\begin{equation}
\label{Hamiltonian3}
\Gamma  = 1/2 P_{\theta }^{2} + (R/2)^{2} \sin^{2}{\theta} P_{t} - R = 0,
\end{equation}
where $P_{\theta} = - (R/2)\sin{\theta} P$ and $F_{3} = -(R/2)P \cos{\theta}$.
The Hamiltonian~(\ref{Hamiltonian3}) is the simple pendulum equation.

That regularization procedure of two centers is similar to Thiele-Burrau
regularization~\cite{VS}. We consider in our regularization an angle variable
$(\theta \in [0:\pi])$ inside of the molecule and a variable $\chi \in
[0:\infty]$ outside of the molecule. The lines $\chi=$ constant are ellipses
and the lines $\theta=$ constant are hyperbola in the physical Cartesian
coordinates plane, similar to the confocal elliptic coordinates~\cite{EH}.

The regularization above is applied to the hydrogen molecule that is related to
coupling of two Kepler problems of two centers.

\subsection*{1-D Hydrogen Molecule}

The hydrogen molecule in 3-D has 6 degrees of freedom (dof) for fixed nuclei.
Since the energy is a constant of movement the system can be reduced to 5 dofs.
The angular moment is not a constant in general. However, the angular moment can
be a constant depending on the initial conditions, e.g., for pendulum periodic
orbits.

The hydrogen molecule can be studied in 1-D as the helium atom~\cite{WRT}. The
helium atom has two configurations in 1-D named Zee (stable orbit - with two
electrons ($e$) in the same side of the nucleus ($Z$)) and $eZe$ (unstable
orbits - with two electrons in each side)~\cite{WRT}. On the other hand, the
H$_{2}$ has one pendulum and four axial configurations.

a) Pendulum trajectories - as defined by Pauli~\cite{WP} for H$^{+}_{2}$ - two
electrons  belong to the middle plane perpendicular to the internuclear axis of
the molecule. The electron oscillation may be symmetric stretch (SS - Wannier
orbit) or  asymmetric stretch (AS).

b) Axial trajectories - the electrons are in the internuclear
axis. Four distinct combinations are
possible because there are two electrons and two protons, which divide
the internuclear axis into three regions.
These four distinct combinations can be named $eZZe$, $ZeeZ$, $ZeZe$, and
$ZZee$ considering the two nuclei ($Z$) and the two electrons ($e$). SS and AS
periodic orbits (PO) can be found for the axial configurations.

The 3-D Hamiltonian is essential to describe the H$_{2}$ system completely. 
However, the 1-D approximation can be used. The difficulties of the
multidimensional calculation are drastically reduced in 1-D model.

The dynamical of the axial trajectories does not have any other constant of
movement than that of the energy. The angular moment is a constant of the
movement for that of the pendulum trajectories. In the last case the whole
system can be reduced to 4 dofs.

The axial and pendulum configurations (1-D) are important parts of the complete
phase space for H$_{2}$ like He atom. There is no place for SS PO in 1-D for
He. However, for H$_{2}$ they are possible since the triple collision is voided,
except to $ZeZe$ configuration. The SS POs can have an important contribution in
high dimension for He and H$_{2}$, e.g., Bohr and Langmuir orbits. Anyway, the
perpendicular motion to orbit must be stable in order to permit the
perpendicular harmonic oscillator quantization since the bounded systems need a
certain classical stability.

The 3-D Hamiltonian and its regularization are shown in reference~\cite{ALC1}.
The 1-D Hamiltonian for H$_{2}$ is given by:

\begin{equation}
\label{Hamiltonian4}
H_{axial} = \frac{(P_{1}^{2}+P_{2}^{2})}{2} - \frac{1}{|z_{1}-R/2|} -
\frac{1}{|z_{1}+R/2|} - \frac{1}{|z_{2}-R/2|} - \frac{1}{|z_{2}+R/2|} +
\frac{1}{|z_{1}-z_{2}|},
\end{equation}
\ \\
where $P_{i}$ is the momentum conjugated at position $z_{i}$ of the $i-$electron
and $R$ is the internuclear parameter. The variables and the parameters can be
calculated for other geometries with a scaling relation as $z=\tilde{z}(R/2),
t=\tilde{t}(R/2)^{3/2}, S=\tilde{S} (R/2)^{1/2}, E=\tilde{E} (2/R)$, etc, where
$\tilde{O}$ is a variable or parameter obtained with $R=2$ and $z, t, S$ and $E$
are the electronic coordinate, time, action and energy, respectively.

The 1-D procedure regularization added a centrifugal barrier over each nucleus
that represents the true barrier in 2-D~\cite{LCAOA}. That barrier implies a
parabolic orbit ($e=1$, where $e$ is the eccentricity) of the electrons around
the nuclei. In this case, the H$_{2}$ system divides the space in three regions
separated by singular potentials. (A smoothed potential (without barrier) will
connect the movements of electrons (with enough energy) in these three regions).
The pendulum configuration is governed by a smoothed potential like to He model
(see below for more details). The regularization is not necessary for pendulum
configuration since $R \neq 0$ and $|z_{1}-z_{2}| \neq 0$ for limited energy.

The Hamiltonians for $ZeeZ$, $ZeZe$, $eZZe$ and $ZZee$ configurations can be 
obtained from Hamiltonian~(\ref{Hamiltonian4}) or from Eq~(\ref{Hamiltonian2})
and Eq~(\ref{Hamiltonian3}) for two-electron molecule.

The Hamiltonian for $ZeeZ$ configuration is obtained with $\chi_{i}=0$ and 
$P_{\chi_{i}}=0$ and it is
\begin{eqnarray*}
\Gamma_{ZeeZ} =  
\frac{1}{2} \frac{\sin^{2}{\theta_{2}}}{(R/2)^{2}}P_{\theta_{1}}^{2} + 
\frac{1}{2} \frac{\sin^{2}{\theta_{1}}}{(R/2)^{2}}P_{\theta_{2}}^{2} - \\
\frac{2\sin^{2}{\theta_{2}}}{(R/2)} -
\frac{2\sin^{2}{\theta_{1}}}{(R/2)} +
\sin^{2}{\theta_{1}}\sin^{2}{\theta_{2}}(P_{t}+\frac{1}{R_{12}}),
\end{eqnarray*}
where $ R_{12}=(R/2)|\cos{\theta_{1}}-\cos{\theta_{2}}|$.

For $ZeZe$ configuration $(\chi_{1}=0, P_{\chi_{1}}=0, \theta_{2}=0 $ and $
P_{\theta_{2}}=0)$ the Hamiltonian is
\begin{eqnarray*}
\Gamma_{ZeZe} = \frac{1}{2}\sin^{2}{\theta_{1}}P_{\chi_{2}}^{2} + 
\frac{1}{2} \frac{\chi_{2}^{2}}
{(R/2)^{2}[\chi_{2}^{2}+(R/2)^{2}]}P_{\theta_{1}^{2}} - \\
\frac{2\chi_{2}^{2}}{(R/2)[\chi_{2}^{2}+(R/2)^{2}]} -
\frac{2\sin^{2}{\theta_{1}}}
{[\chi_{2}^{2}+(R/2)^{2}]^{1/2}} +
\frac{\sin^{2}{\theta_{1}}\chi_{2}^{2}}
{[\chi_{2}^{2}+(R/2)^{2}]}(P_{t}+\frac{1}{R_{12}}),
\end{eqnarray*}
where $ R_{12}=[\chi_{2}^{2}+(R/2)^{2}(\cos^{2}{\theta_{1}}+1) - R
\sqrt{\chi_{2}^{2}+(R/2)^{2}} \cos{\theta_{1}}]^{1/2}$.

For $eZZe$ configuration $[\theta_{1}=\pi(\cos{\theta_{1}}=-1), 
\theta_{2}=0(\cos{\theta_{2}}=1) $ and $ P_{\theta_{i}}=0]$ 
and for $ZZee$ one $(\theta_{i}=0 $ and $ P_{\theta_{i}}=0)$ 
the Hamiltonians are similar, differing by $R_{12}$
\begin{eqnarray*}
\Gamma_{eZZe} = \Gamma_{ZZee} = \frac{1}{2} \frac{\chi_{2}^{2}}
{[\chi_{2}^{2}+(R/2)^{2}]}P_{\chi_{1}}^{2} + \frac{1}{2}
 \frac{\chi_{1}^{2}}
{[\chi_{1}^{2}+(R/2)^{2}]}P_{\chi_{2}}^{2} - \\
\frac{2\chi_{2}^{2}}
{[\chi_{1}^{2}+(R/2)^{2}]^{1/2}[\chi_{2}^{2}+(R/2)^{2}]} -
\frac{2\chi_{1}^{2}}
{[\chi_{1}^{2}+(R/2)^{2}][\chi_{2}^{2}+(R/2)^{2}]^{1/2}} + \\
\frac{\chi_{1}^{2}\chi_{2}^{2}}
{[\chi_{1}^{2}+(R/2)^{2}][\chi_{2}^{2}+(R/2)^{2}]}(P_{t}+\frac{1}{R_{12}}),
\end{eqnarray*}
where $ R_{12}=\sqrt{\chi_{1}^{2}+(R/2)^{2}}+\sqrt{\chi_{2}^{2}+(R/2)^{2}}$ for
$eZZe$ configuration and 
$R_{12}=|\sqrt{\chi_{1}^{2}+(R/2)^{2}}-\sqrt{\chi_{2}^{2}+(R/2)^{2}}|$ for
$ZZee$ one.

The Hamiltonian for pendulum configuration is
$$
H_{pend} = \frac{1}{2}P_{y_{1}}^{2} + \frac{1}{2}P_{y_{2}}^{2}-
\frac{2}{\sqrt{y_{1}^{2}+(R/2)^{2}}}-\frac{2}{\sqrt{y_{2}^{2}+(R/2)^{2}}}+
\frac{1}{[R_{12}]},$$
where $ R_{12}=|y_{1}-y_{2}|$ with $y-$axis is perpendicular to internuclear
axis or $z-$axis.

The last equation is similar to He atom with smoothed
potential~\cite{LCAOA,JHE}, where  $(R/2)$ represents the smoothed parameter
$(\delta=R/2)$, noting that the interelectronic repulsion term is not smoothed.
Therefore, the understanding of the smoothed He system can help to know the
pendulum configuration behavior for H$_{2}$ molecule.

\subsection*{Semiclassical quantization}

Integrable and quasi-integrable systems can be studied with a generalized
Bohr-Sommerfeld procedure or Einstein-Brillouin-Keller (EBK)
quantization~\cite{MG,AMOA}. The EBK quantization works well if the regular
dynamics dominates a system, e.g., systems with weak correlation movements.

The quantization of the action of the periodic orbits is the first step in order
to apply the EBK quantization. The actions of the orbits can be calculated as
\begin{displaymath}
S= 2 \pi (n+\alpha/4)
\end{displaymath}
where $\alpha$ is the Maslov index where for simple rotation $\alpha=0$ and for
simple  libration $\alpha=2$. Nevertheless, it is necessary to modify the
actions in more degree of freedoms as
\begin{displaymath}
S= \sum_{i} 2 \pi [n+\alpha/4+\theta_{i}(l+1/2)]
\end{displaymath}
where $\theta_{i}$ is the stability angle for $i$ degree of freedom belongs to
perpendicular displacements of the orbit. The term $(l+1/2)$ is the properly
eigenvalue of a perpendicular harmonic oscillator

The rigorous EBK quantization of invariant KAM tori around 1-D periodic orbits
is not  applied to the H$_{2}$ system since they are unstable and consequently
the system can not be reduced to  an one-periodic-orbit model. However, the
semiclassical path-integral quantization of nonintegrable Hamiltonian
systems~\cite{MG} can be applied.

A better semiclassical description can be obtained using the Gutzwiller trace
formula~\cite{MG} to calculate the density of levels.

Non-integrable systems are more difficult to study. We need infinity periodic
orbits (PO) to describe a complex system completely. We need to find procedures
of the truncation of the sum (resummation) for practical propose in order to
converge convincing results with finite PO (similarly to use a finite base in
quantum calculation). Each PO can contain implicitly correlation of the movement
similarly to a correlated base (e.g., Hylleras functions) of the quantum
calculation.

The extension of the quantization for most part of the non-integrable systems is
unknown yet. However, the global quantization with classical periodic orbits is
an alternative.

The quantization for a non-integrable system was made by Gutzwiller~\cite{MG}.
The semiclassical density of states (trace formula) is~\cite{WR}

\begin{equation}
\label{trace}
d(E) = \sum_{r} \frac{T_{r}}{\pi \hbar}
\sum_{j \neq 0} \frac{ \cos [ j (S_{r} / \hbar  - \alpha_r \pi / 2) ] }
{|[det({\bf M}_{r}^{j} - {\bf 1})]^{1/2}|} .
\end{equation}
The trace formula above sums all repetitions ($j$) of the primitive periodic
orbits ($r$) with period $T_{r}$, action $S_{r}$ and Maslov index $\alpha_{r}$
and ${\bf M_{r}}$ is the stability matrix. The determinant of the trace formula
depends on the fixed point. In two degrees of freedom, it is given as

\begin{eqnarray*} 
(1/2)[det({\bf M}_{r}^{j} - {\bf 1})]^{1/2} = \left\{ \begin{array}{lll}
-i \sinh (j \lambda /2),    & \mbox{hyperbolic,} \\
\cosh (j \lambda /2), & \mbox{inverse hyperbolic,} \\
\sin (j \pi \theta),     & \mbox{elliptic,}
\end{array} \right.
\end{eqnarray*}
where $\lambda$ is the Lyapunov exponent (real eigenvalue) and $\theta$ is the
stability angle (imaginary eigenvalue), if the system has more than two degrees
of freedom complex (or loxodromic) eigenvalue can be obtained. The singular
spectrum is obtained with all repetitions of the (infinity) primitive orbits in
the Gutzwiller formula. The spectrum is smoothed for a finite number of the
repetitions, e.g., fixing a maximum period $(T_{max})$. The trace formula has
convergence problems, i.e., the sum diverges. Several types of the resummation
were developed~\cite{MG, AMOA} in order to try convergences in the trace
formula. For example, the zeta function representation of the trace formula
works well for semiclassical quantization~\cite{WRT,CE,PC}.

The Gutzwiller trace formula in product representation or Gutzwiller-Voros zeta
function~\cite{WRT,CE} given below.

\begin{equation}
\label{zeta}
Z(E) = \prod_n (E-E_n) \sim \prod_{m_k} \zeta^{-1} = \prod_{m_k} \prod_{p}
(1-t_p^{m_k}),
\end{equation}
where $\zeta^{-1}$ is the dynamical zeta function or simply zeta function, the
weight $t_p$ for each primitive (nonrepeating) periodic orbit (PPO) is given by

\begin{equation}
\label{weightE}
t_{p}^{m_k}(E) = (\pm 1)^{m_k} \exp[2 \pi i (S(E) - \alpha/4) - \sum_k (m_k+1/2)
\chi_k(E)], \end{equation}

or

\begin{equation}
\label{weight}
t_{p}^{m_k}(z) = (\pm 1)^{m_k} \exp[2 \pi i (z \tilde{S}(Rez) - \alpha/4) - \sum_k
(m_k+1/2) \chi_k(Rez)], \end{equation}
where $z$ is the effective quantum (complex) number with real part given by
$Rez=(-E)^{-1/2}$. $\tilde{S}(Rez)=S(E)(-E)^{1/2}=S(E)/Rez$, where $S(E)$ is the
action for energy $E$, $\alpha$ is Maslov index and $\chi_k$ is the stability
exponent associated to eigenvalue $k$ of the stability matrix for each PPO that
depends on the energy for this non-scaled energy system. Note that the
$\tilde{S}$ is not a scaled action since the H$_2$ is not scaled system.
$\tilde{S}$ is a convenient notation well defined in the real axis. $m_k$ gives
the bending excitation, we considered $m_k=0$.

The symmetry decomposition of the zeta function is obtained considering that the
collinear hydrogen molecule has a $C_2$ symmetry. The symmetric subspace of the
zeta function ($\zeta_{+}^{-1}$) is obtained maintaining the $A_1$ symmetry of
the $C_2$ group. The antisymmetric subspace of the zeta function
($\zeta_{-}^{-1}$) is obtained with $A_2$ symmetry. The $\zeta_{+}^{-1}$ is
correlated to symmetric quantum resonances (singlet states) and the
$\zeta_{-}^{-1}$ with antisymmetric quantum resonances (triplet states).

The POs are classified in boundary (denoted by $+$), symmetric ($-$) and
non-symmetric ($NS$). The zeta function gives the resonances of the PO The
factorized form of the zeta function for non-desymmetrized PPO is given
by~\cite{CE} for each 1-D configuration

\begin{equation}
\label{zetaM}
\zeta_{+}^{-1} = (1-t_{+})(1-t_{-})(1-t_{NS})
\end{equation}
and
\begin{equation}
\label{zetam}
\zeta_{-}^{-1} = (1+t_{-})(1-t_{NS})
\end{equation}
where $t_{+}$, $t_{-}$ and $t_{NS}$ are the weight for boundary, symmetric and
non-symmetric PPOs. The zeta function in symmetry factorized form is given by
\begin{equation}
\label{zetat}
\zeta^{-1} = \zeta_{+}^{-1} \zeta_{-}^{-1}
\end{equation}

Each 1-D configuration of the H$_2$ is described by a Eq~(\ref{zetat}). We build
a new zeta function given by
\begin{equation}
\label{zetaconf}
\zeta^{-1} = \zeta_{ZeeZ}^{-1} \otimes \zeta_{eZZe}^{-1} \otimes
\zeta_{ZZee}^{-1} \otimes \zeta_{pendulum}^{-1}
\end{equation}
where the contribution for quantum resonances for each configuration is
considered in a same zeta function. The zeta function~(\ref{zetaconf}) can be
factorized, as in Eq~(\ref{zetat}), in order to obtain the respective symmetric
and antisymmetric zeta functions for all configurations. The $\otimes$ operation
can be direct product or direct sum depending on the approximation, i.e., if the
configurations are not or are approximately independent to each other.

\subsection*{Quantum Methods}

Several methods can be used in order to calculate the Schr\"odinger equation in
the adiabatic approximation (fixed nuclei) for excited states of the 
two-electron systems~\cite{TRR}:

a) A Hartree-Fock calculation as an initial guess to obtain the molecular base 
following an electronic correlation correction with Configuration Interaction (CI)
or Perturbation Theory (PT);

b) The selection of important contributions obtained via CI can be made with
Feshbach projection operator in order to describe the resonance states;

c) Independent particle method with following a PT in order to correct the
interelectronic repulsion term obtained via C.Neumann formula~\cite{YS};

d) R matrix method;

e) Complex rotation method, etc.

Those methods can be applied suitable depending on the behavior of the system or
states.

The calculation of the Schr\"odinger equation is not difficult to realize for
quantum systems when the correspondent classical ones are integrable or
quasi-integrable, i.e., one-electron systems or many-electrons ones with low
degree of the correlation. An initial guess base can give an excellent numerical
convergence for those systems. The difficulties in the convergence increase with
correlation movement and with quantum number.

There are serious difficulties to find a good guess base for correspondent
non-integrable systems since a base is a (approximative) solution of an
integrable system. Because of that the numerical convergence is very hard to
manage for non-integrable systems. A good strategy is to find, if exists, a
particular coordinate which can describe explicitly the quasi separability of
the problem. For example, the adiabatic quantum calculation of the frozen
planetary atom~\cite{WRT}.

\subsection*{Quantum $X$ Semiclassical}

We can summarize some important aspects of the quantum and semiclassical
calculations.

Advantages of quantum calculation:

a) The quantum theory is correct;

b) It is possible to calculate any system via brute force;

c) Very known theory.

Disadvantages of the quantum calculation:

a) Difficulties to interpret the results;

b) Difficulties to treat highly excited states;

c) High computational costs.

Advantages of the semiclassical calculation:

a) Suitable on interpretation of the results;

b) Suitable to treat highly excited states;

c) Low computational costs.

Disadvantages of the semiclassical calculation:

a) The semiclassical theory is approximate with few exceptions;

b) It is hard to find the solution for generic systems;

c) Unknowing of the generic recipes to quantization of the chaotic systems.

\section{Results and Discussions}

Our starting point of the studies of the H$_{2}$ is to compare this two-electron
molecular system to a two-electron atomic, i.e., the helium atom. The He atom
was studied semiclassically~\cite{LP,WRT,TRR,SG} and very important results are
managed.

\subsection*{Classical studies}

Firstly, we have to integrate the Hamilton equations from H$_{2}$ numerically
with a variable  step fourth-order Runge-Kutta method in order to obtain the
periodic orbits (PO). Some those PO are shown in Fig.1 and Fig.2 for $R=2.0$au.

The Fig.1 shows the potential energy level curves for axial configurations with
a smoothed parameter $\delta=1$ in order to avoid the divergences in level
curves. Of course, we do not consider the smoothed parameter in our
calculations. The abscissa and ordinate axes represent each other one electron.
The protons are located at $(+1,-1)$ and $(-1,+1)$ (absolute minima of the
potential energy),where all axial POs emerge, exception for $ZZee$
configuration. The saddle point is located at origin (0,0). The diagonal line
$(x=y)$ is the place where the electronic repulsion is infinity. The energy of
the POs increases with their size for a same configuration.

The $eZZe$ Wannier (SS) and fundamental (AS) POs in Cartesian coordinates for
several  values of energy $(E=-2.5$au (SS and AS), $-2.0$ and $-1.0)$ are shown
in the places on the potential energy curves labeled with $eZZe$ in Fig.1. The
fundamental orbit is an AS PO where the period is the shortest of those orbits
belonging to same AS configuration. That orbit is similar to fundamental PO of
the $eZe$ configuration of the He~\cite{WRT}. The $ZeeZ$ fundamental POs in
Cartesian coordinates for $E=-2.41$au (SS and AS), $-2.0$ and $-0.55$ are shown
in $ZeeZ$ place. The $ZeZe$ fundamental AS POs in Cartesian coordinates for
$E=-3.71$au, $-2.66$ and $-2.4$ are shown in $ZeZe$ and $eZeZ$ places. Near SS
POs (simultaneous double collision) for $ZeZe$ at $(+1,-1)$ and for $eZeZ$ at
$(-1,+1)$ and near SS POs ( triple collision) for $ZeZe$ at $(+1,+1)$ and for
$eZeZ$ at $(-1,-1)$ are not considered. The molecular Frozen Planetary
Configuration (FPC)~\cite{ALC1} POs of the ZZee configuration in Cartesian
coordinates for $E=-5.65$au and $-4.01$ are shown in $ZZee$ and $eeZZ$ places.
Those POs received the molecular FPC name by analogy with Frozen Planetary Atom
for He atom~\cite{WRT}.

\begin{figure}
\begin{center}
\vskip 10pt
\includegraphics[width=15cm, height=15cm]{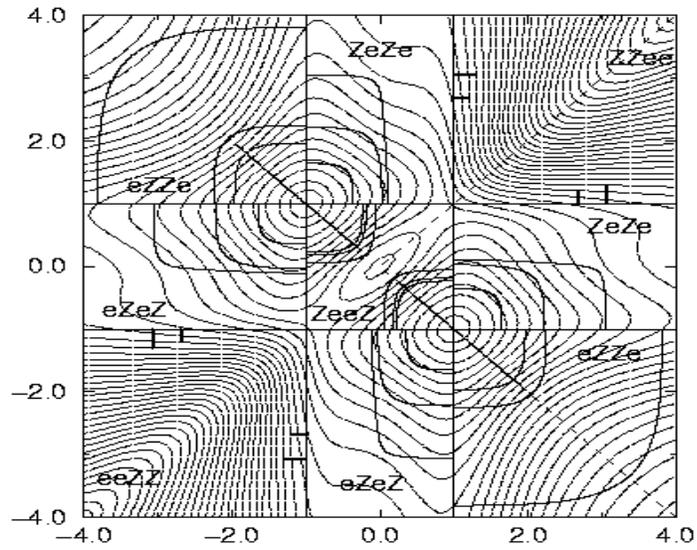}
\caption{Potential energy level curves for axial configurations with smoothed
parameter $\delta=1$ and $R=2$. Each axis represents one electron coordinate.
The protons are located at $(+1,-1)$ and $(-1,+1)$. The Wannier and fundamental
POs in Cartesian coordinates for several values of energy are put in their
respective configurations. The dashed lines shows the desymmetrized region.}
\end{center}
\end{figure}

\begin{figure}
\begin{center}
\vskip 10pt
\includegraphics[width=15cm, height=15cm]{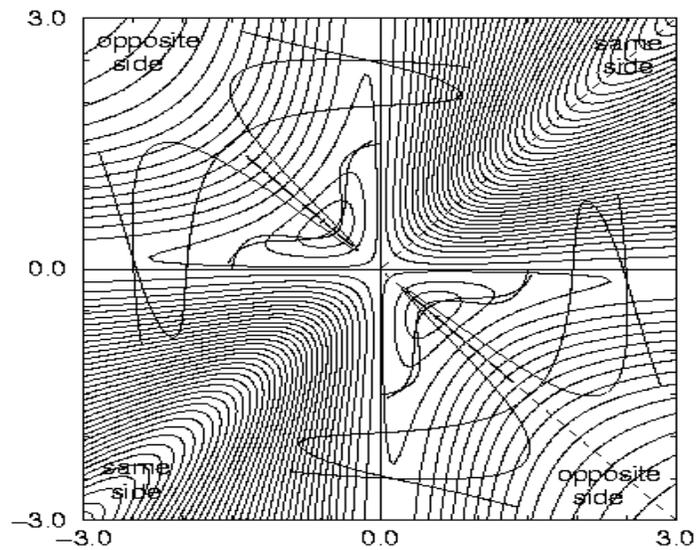}
\caption{Potential energy level curves for pendulum configuration with
smoothed parameter $\delta=1$ and $R=2$. Each axis represents one
electron coordinate.The pendulum orbits born in $3^{-1/2}(\pm 1,\mp 1)$.
The dashed lines shows the desymmetrized region.}
\end{center}
\end{figure}

The Fig.2 shows the potential energy level curves for pendulum configuration.
Each axis represents one electron coordinate. The potential energy level curves
show two absolute minima of the energy, where all pendulum POs emerge. The
diagonal line $(x=y)$ is where the electronic repulsion takes place. The
diagonal $(x=-y)$ lines are the Wannier POs in Cartesian coordinates for
$E=-2.48$au located in the $opposite$ $side$ labeled place. The pendulum
fundamental POs in Cartesian coordinates for $E=-2.5979$au (near to minima),
$-2.47, -2.45$ (energy bifurcation), $-1.60$ and $-1.60$ (two different orbits
with similar energy) are shown in $opposite$ $side$ place for $E < -2.45$. The
POs extend to $same$ $side$ for $E > -2.45$, i.e., after bifurcation. The energy
of the POs increases with their size.

We calculated the stability exponent of the ``fundamental'' orbits of all axial
and pendulum configurations. We also obtained the stability indexes of the
Wannier orbit for $eZZe$, $ZeeZ$ and pendulum configurations. The stability
angles and the Lyapunov exponents of the periodic orbits~\cite{MG,AMOA} were
obtained by the Monodromy method~\cite{BD,NSS}.

\subsubsection*{Stability exponents}

The He is scaled with energy, but H$_2$ with $R$ parameter (internuclear
distance) is not  scaled with energy. It is necessary to calculate the stability
index for all range of energy $(E)$ parameter where each point can be scaled
with $R$. (The parameters $R$ and $E$ can be exchanged in relation to scaling
law.) The Monodromy matrix of He atom in 2-D has two trivial pairs of
eigenvalues. However, H$_2$ molecule in 2-D has only one trivial pair of
eigenvalues for axial configurations. The angular momentum is not a constant of
the movement for H$_{2}$ molecule since the symmetry of that is less than of the
He atom. The motion belonging to axial orbit for H$_2$ gives one trivial pair of
eigenvalues of the monodromy matrix and other generally real ones (hyperbolic).
The perpendicular motion of the 1-D orbit gives two non-trivial pairs of
eigenvalues and the He atom gives one non-trivial pair of eigenvalues.

The Wannier orbits has infinity Lyapunov~\cite{WRT} for He atom. However, the
Wannier orbit for axial and pendulum configurations have finite Lyapunov. The
Wannier orbit for pendulum configuration coincides with Bohr (BH$_{2}a$) and
Langmuir (LH$_{2}a$) orbits~\cite{ALC} for parabolic eccentricity  ($e=1$). The
Wannier orbits for $ZeeZ$ and $eZZe$ configurations coincide with the 1-D
projection of the BH$_{2}b$ and LH$_{2}b$ orbits~\cite{ALC}. The 1-D Wannier
orbits can contain contribution of those orbits in higher dimension~\cite{ALC1}.

The effective potential of the pendulum Wannier orbit is discussed below. We can
compare the H$_{2}$ system to the He atom. The symmetric potential function of
the He is

\begin{equation}
\label{pot1}
V_{He}(z) = -7/(2z),
\end{equation}
where $z$ is the module of the electronic position vector. The
potential~(\ref{pot1}) is an assympthotical approximation for triple collision.

The Wannier orbit potential of the pendulum configuration is

\begin{equation}
\label{pot2}
V_{H_{2}}(z) \approx V_{He}(y) +R^{2}/(2y^{3}).
\end{equation}
The potential~(\ref{pot2}) presents a minimum that stabilizes the system.

Now we discuss the stability indexes in order to initiate the classical studies
of the hydrogen molecule. The stability index curves are shown in Figs.3a to 3h
for all 1-D PO calculated for H$_{2}$ with $R=2$a.u., all solid lines are the
stability angles and the dashed lines are the Lyapunov exponents.

The stability indexes of the H$_{2}$ differ from He by an important symmetry
property discussed ahead. The direct consequence is: the H$_{2}$ molecule has a
bigger number of pairs of the non-trivial stability matrix eigenvalue [see
Eq.~(\ref{trace})] than that for He. For example, the number of the non-trivial
pairs of the eigenvalues or equivalently, stability indexes for axial
configuration(s) of the He (H$_{2}$) with four dof's is two (three): one
correspondent to the parallel dof to PO and one (two) perpendicular one(s).
Following we discuss the stability indexes of the H$_{2}$.

The stability indexes of the Wannier and fundamental orbits of the $eZZe$
configuration are shown in Figs.3a and 3b, respectively for $-2.0<E<-0.33$. The
Fig.3a has two curves, the curves are the stability angle divided by $2 \pi$ for
perpendicular (stable) dof ($y-$axis). The curves are the same for $x-$axis
since the system has azimuthal symmetry. The lowest curve describes a stable
behavior until $E \approx -0.65$au, besides the behavior change to unstable and
the curve becomes the Lyapunov divided by $2 \pi$. The Fig.3b has three curves,
the lowest curve is the Lyapunov divided by $2 \pi$ for parallel (unstable) dof
of the PO. The other curves are the stability angle divided by $2 \pi$ for
perpendicular (stable) dof ($y-$axis). The lowest curve describes a stable
behavior until $E \approx -0.4$au. Besides the behavior becomes unstable. The
curves for $E<-2.0$ follow same behavior and the Lyapunov curves are multiplied
by 10 for Figs.3a and 3b.

The eigenvalue of the stability matrix for axial dof for PO of the $ZeeZ$,
$eZZe$ and $ZeZe$ configurations are real, i.e., they describe an unstable
behavior similar to $eZe$ of the He. The eigenvalues of the perpendicular dof
are imaginary (elliptic) or real (hyperbolic) depending on the energy range.
That behavior is different from He since the perpendicular dof is ever
imaginary. The pendulum configuration shows similar behavior of the axial ones.

\begin{figure}
\begin{center}
\vskip 10pt
\includegraphics[width=15cm, height=15cm]{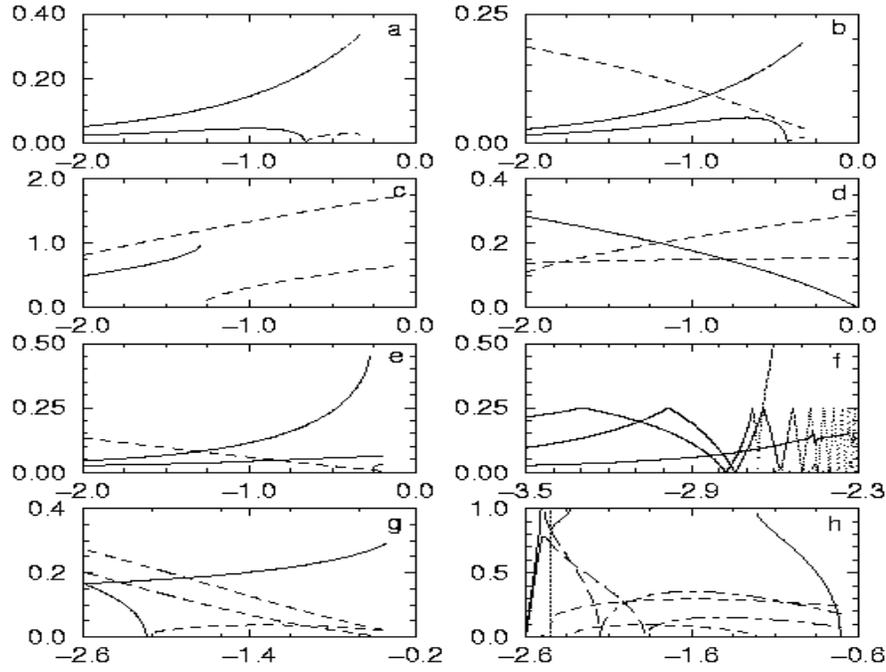}
\caption{The stability index curves for 1-D POs with $R=2$au as a function of
electronic energy for $R=2$: (a) Wannier-eZZe; (b) eZZe; (c) Wannier-ZeeZ; (d)
ZeeZ; (e) ZZee; (f) ZeZe; (g) Wannier-pendulum, and (h) pendulum orbits (1) and
(2). The continuous lines are the stability angles and the dashed ones are the
Lyapunov exponents. The stability angles for second pendulum PO (2) are given by
long-dashed curves and the Lyapunov exponents by dot-dashed curves for same PO
in figure h. The Lyapunov curves are multiplied by 10 in figures a, b, e and h}
\end{center}
\end{figure}

The stability indexes of the Wannier and fundamental orbits of the $ZeeZ$
configuration are shown in Figs.3c and 3.d, respectively for $-2.0<E<-0.0$. The
Fig.3c shows the Lyapunov and the stability angle are for perpendicular dof. The
Fig.3d shows the Lyapunov curve for parallel dof as a lowest curve. The other
Lyapunov and the stability angle curves are for perpendicular dof. These last
stability indexes are components of a complex (loxodromic) eigenvalue. However,
those indexes change to two imaginary eigenvalues (stability angles) for
$E<-2.33$. The action and the period do not increase with energy for those POs
since the electrons are limited by two singular potential for ZeeZ
configuration. That behavior is distinct from any other 1-D PO.

The stability indexes of the molecular FPC PO of the $ZZee$ configuration are
shown in Fig.3e for $-2.0<E<-0.2$. The intermediary curve is the stability angle
for parallel dof. The highest curve is the stability angle for perpendicular dof
and the lowest one is the Lyapunov for same perpendicular dof. The curves for
$E<-2.0$ follow similar behavior. The Lyapunov curves are multiplied by 10.

The PO of the $ZZee$ configuration (molecular FPC) in 1-D is stable as FPA for
He~\cite{WRT}, but in 2-D the molecular FPC is unstable since the inner electron
can not shield the farthest proton and the outer electron can be attracted by
that proton unstabilizing the system. The stability index in axial direction
shows stability for all range of energy, the perpendicular dof changes the
stability depending on the energy range.

The stability indexes of the fundamental orbit of the $ZeZe$ configuration are
shown in Fig.3f for $-3.5<E<-2.3$. The curves are the stability angle for all
dof with some oscillations between the stable and unstable behavior. That PO is
totally stable for $E < -2.65$au, it is the only case since known of the
non-ergodicity of H$_2$ system. All exponent indexes are represented by dotted
lines in the Fig.3f. The stability angles are limited to 0.25, but the Lyapunov
exponent is not. The highest curve belongs to the parallel dof for $E = -3.5$.

The stability indexes of Wannier orbit of the pendulum configuration are shown
in Fig.3g for $-2.6<E<-0.4$. That figure has four curves, two dashed curves are
the Lyapunov for perpendicular dof in axial direction. The highest continuous
curve is the stability angle for other perpendicular dof and the lowest one is
the stability angle for $E<-2.17$ for increasing $E$ is the Lyapunov for
parallel dof.

The stability indexes of the two fundamental orbits of the pendulum
configuration are shown in Fig.3h for $-2.6<E<-0.7$. That figure has four
curves, two curves for each PO. Similar to Wannier PO, the fundamental PO for
pendulum configuration has four non-trivial stability matrix eigenvalues.
However, the stability exponents (Lyapunov) for perpendicular dof in axial
direction are not shown. The stability indexes are the same for the two PO with
$E<-2.45$au. Those indexes become different for each PO when $E>-2.45$au. A
bifurcation appears for $E \approx -2.45$au and emerge two orbits approximately
with equal periods. The vertical dotted line in Fig.3h separates the energy
regions "before" and "after" bifurcation (see discussions below). The four
indexes oscillated between the stability angle and Lyapunov exponents. The
stability angles for second PO are given by long-dashed curves and the Lyapunov
exponents by dot-dashed curves. The Lyapunov curves for the two pendulum POs are
multiplied by 10.

The two dofs perpendicular to the POs for 1-D pendulum configuration are not
equivalent that is different from axial configuration and for 1-D He atom. The
plane that contains 1-D pendulum POs and the perpendicular dof, non-parallel to
internuclear axis, has rotational symmetry.

Some simple calculations can be performed in order to study the behavior of the
pendulum  configuration.

\subsubsection*{Pendulum configuration behavior}

The absolute minimum of energy is obtained considering the electrostatic
equilibrium for $y_1=-y_2$ with energy $E=-3\sqrt{3}/R$ and $y_1=R/(2\sqrt{3})$.
Below of that energy is not classically permitted for pendulum configuration.
The electrons remain fixed in the minimum. All orbits  start of that minimum for
pendulum configuration (see Fig.2).

The Hamiltonian of the pendulum configuration is very near to the He atom with
smoothed potential~\cite{LCAOA}. The internuclear dof is always unstable, the He
analogy is valid  for perpendicular plane of the internuclear axis. The
fundamental pendulum orbit is marginally stable (integrable) for
$E=-3\sqrt{3}/R$ and stable for the range $-3\sqrt{3}/R < E < E_{c}$, where
$E_{c}=-(2/R)(2+(2^{2/3}-1)^{3/2})$ is a critical energy. The orbit is similar
to $eZe$ configuration of the He for stable region, i.e., the electrons are in
opposite side of the He with smoothed potential (internuclear axis). Note that
the He orbit is stable because of the smooth parameter $(R/2)$. The fundamental
PO connects the ``$eZe$'' (opposite side) configuration with ``$Zee$'' one (same
side) for $E > -(2/R)(2+(2^{2/3}-1)^{3/2})$, in this energy the system becomes
unstable for increasing energy.

The $E_{c} = -(2/R)(2+(2^{2/3}-1)^{3/2})$ is found considering the electrons
remain stopped (kinetic energy is null) in, e.g., $y_1=0$ and
$y_2=-(R/2)(2^{2/3}-1)^{-1/2}$, for increasing energy $(y_1<0)$ the system
connects the $eZe$ and $Zee$ smoothed configurations. FPC orbit is not possible
for smoothed He with ``$Zee$'' configuration since ``$Zee$'' is ever connected
to hyperbolic ``$eZe$'' configuration.

The Fig.3h shows the stability exponents for two pendulum orbits, which
bifurcated in transition $eZe ---> eZe + Zee$ with $E=-2.45$ and $R=2$ (see
Fig.2).

\subsubsection*{Axial Wannier orbits}

The axial Wannier orbits for $ZeeZ$ and $eZZe$ configurations present
simultaneous binary collisions (SBC). When the two electrons collide
simultaneously with the protons the regularization function~Eq.(\ref{regfunc})
for two-electron system approaches from zero as square potential and the
Hamiltonians $\Gamma_{ZeeZ}$ and $\Gamma_{eZZe}$ diverge. It is necessary to
change the reguralization strategy. The treatment of the SBC is known~\cite{CS}.
However, we obtained the eigenvalues belonging to the perpendicular degree of
freedom (dof) of the Wannier orbits for ZeeZ and eZZe configurations by an
alternative simple way.

The principal argument is based on symmetry reductions:

1) The full Hamiltonian is reduced to 2 dof for an effective electron
interacting with its image, i.e., $\chi_1 =-\chi_2$ and $\theta_2 =\pi
-\theta_1$. The effective Hamiltonian is very easy to calculate and the
stability exponents are: one trivial (energy conservation) and other non-trivial
correspondent to perpendicular dof.

Following, the stability exponents are obtained with Hamiltonians with 4 dofs.

2) It was considered the coupling of the two H$_2^+$ systems, where the
interaction term $(R_{12}^{-1})$ and the energy were multiplied by the factor
$(f_1 + f_2)$ instead $(f_1 * f_2)$~\cite{ALC1}, where $f_i$ is the
regularization function for electron $i$. The exact Hamiltonian is recovered in
the limit of the symmetric stretch. The stability exponents are correct for
perpendicular dof, but one eigenvalue for parallel dof is not correct.

The following approximations confirm those results.

3) The factor $(f_1 + f_2)$ was substituted by $(f_1 * f_2)^{1/2}$ and same
eigenvalues for perpendicular displacement are obtained.

Those results are already confirmed with the last approximation:

4) The complete Hamiltonian in 4 dofs (without any approximations)~\cite{ALC1}
was considered. However, the neighborhoods of the Wannier orbits with dephased
the collisions were the starting trajectories. The calculation with this
quasi-orbit is very difficult since the neighborhoods are chosen very near to
the Wannier orbit to avoid significant numerical deviation of the previous
calculations. The same eigenvalues for perpendicular dof are obtained, but the
non-trivial eigenvalue parallel to the internuclear axis can not be obtained
satisfactorily, although it does not matter since the most important eigenvalues
for semiclassical quantization are the perpendicular ones. A more elaborated
calculation~\cite{CS} is necessary to obtain all eigenvalues.

The instability related to the parallel dof of the internuclear axis ($z$-axis)
is high in comparison to other axis for axial and pendulum configurations.
Because of that the semiclassical quantum contribution of that dof is little.
Therefore, the dynamic motion in the plane perpendicular to internuclear
(bonding) axis is the most important in the semiclassical quantization (see
above).

The $eZe$ configuration is hyperbolic and the $Zee$ one is near to integrable
for He atom~\cite{WRT} and for H$^{-}$ ion~\cite{GR}. The H atom is integrable
system. The H$_2$ molecule has a intermediary behavior between the H and He
atoms, i.e., between the near integrable and hyperbolic behaviors. The
correlation diagram among the configurations of the H$_2$ molecule with atoms is
showed in Fig.4

\begin{figure}
\begin{center}
\vskip 10pt
\includegraphics[width=15cm, height=15cm]{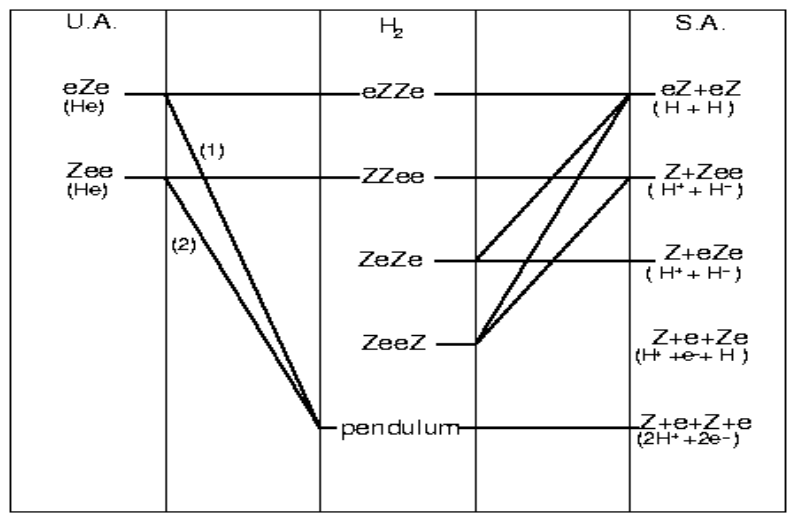}
\caption{Correlation diagram among the configurations of the H$_2$ molecule with
the H$^+$, \, H, \, H$^-$ (separated atoms - SA), and He atoms (united atom -
UA). The correlations labeled with (1) and (2) appear for E$>$E$_c$ for pendulum
configuration and for E$<$E$_c$ only appears the correlation (1)}
\end{center}
\end{figure}

\subsection*{Semiclassical studies}

\subsubsection*{Single Quantization}

We obtained some results of the single quantization of the 1-D orbits. We
considered the following action
$S(n,E;R)=2\pi[n+\alpha/4+(1/2)\sum_{i=1}^{N} \theta_{i}(E)/2\pi)]$, where $n$
is the principal quantum number, $\alpha$ is the Maslov indices ($\alpha=4$ for
all 1-D orbits considered), $\theta_{i}(E)$ is the stability angle and $N$ is a
integer number that depends on the orbit, the energy and the dof and $R$ is a
parameter. The axial configurations have degenerated (isotropic) perpendicular
directions, i.e., the stability indexes are equal for the two direction
perpendicular to bonding distance. The all directions of the pendulum
configuration are different.

Curves of single quantization for some 1-D orbits are shown in Fig.5. We used
the scale law in order to obtain the total energy (with nuclear repulsion term)
as a function of internuclear distance (R). The lower curve with circle is the
ground state of the hydrogen molecule. The single quantization results are: the
lines labeled with {\bf w} are for Wannier orbits; the lines labeled with {\bf
p} are for pendulum orbits; the lines with {\bf a1} are for $eZZe$ orbit and the
lines with {\bf a2} are for $ZeeZ$ orbit.

\begin{figure}
\begin{center}
\vskip 10pt
\includegraphics[width=15cm, height=15cm]{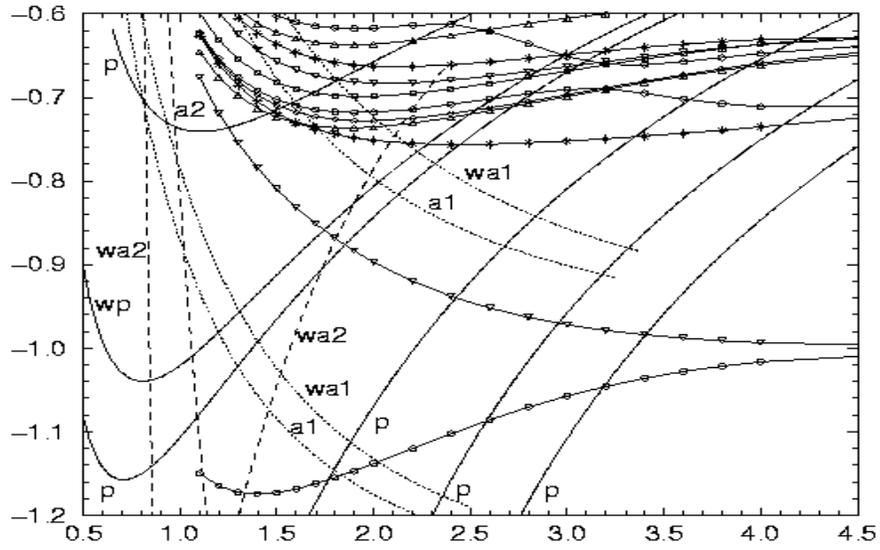}
\caption{The total energy as a function of internuclear distance (R). The
quantum states are labeled with symbols: $\circ - ^1\Sigma_{g}^{+}$;
$\bigtriangledown - ^3\Sigma_{u}^{-}$; $\ast - ^1\Sigma_{u}^{-}$;
$\bigtriangleup - ^3\Sigma_{g}^{+}$; $\diamond - ^3\Pi_{g}$ and (square) $^1\Pi_{g}$. The single quantization results are: the solid line labeled with
{\bf wp} is for Wannier pendulum orbit; the solid lines labeled with {\bf p} are
for pendulum orbits; the dotted lines with {\bf wa1} and {\bf a1} are for $eZZe$
Wannier and fundamental orbits and the dashed lines with {\bf wa2} and {\bf a2}
are for $ZeeZ$ Wannier and fundamental orbits, respectively}
\end{center}
\end{figure}

The single quantization does not work well to describe the hydrogen molecule for
intermediary value of the bonding distance (R). Since the molecular quantum
states studied are composed approximately by a linear combinations of the atomic
quantum states or, in semiclassical words, by a combination of the POs of the
atomic limits ($R \rightarrow 0$ and $R \rightarrow \infty$). The single
quantization shows that all POs can contribute in a description of the quantum
states. The $eZZe$ POs can be the most important ones as the asymmetric stretch
PO of the $eZe$ configuration of the He atom.

We show that the single quantization can not describe the first states of
the H$_2$ system and a more elaborated theory can be used in order to account a
better result. However, since the H$_2$ system is not a hyperbolic system as He
atom, it is possible to find some stability island which are suitable for WKB
quantization, e.g., for pendulum orbits around the absolute minimum of energy
($E=-3 \sqrt{3}/R$, see Fig.2) and for $ZZee$ and $ZeZe$ orbits (see Figs3.e and
3.f). Those WKB quantization will be connected to particular subspectrum of the
H$_2$ system.

\subsubsection*{Semiclassical Quantization}

The density of levels was obtained from trace formula with few repetitions,
i.e., fixing a maximum period $(T_{max})$. The $T_{max}$ was considered as the
period of the fundamental pendulum orbits $(T_{pend})$ since this period
is longer than the others considered here. The orbits are summed several times
depending on the ratio $T_{i}/T_{pend}$, where $T_{i}$ is the period of the orbit
considered. The ratio $T_{i}/T_{pend}$ changes according to energy, since the
system is not scaled to energy with $R$ fixed. The number of the repetitions is
Int$[T_{i}/T_{pend}]$, where Int$[X]$ is the entire part of the $X$.

We study the possibility to include the axial and pendulum orbits (6-dof) in the
same trace formula. The axial orbits have five eigenvalue pairs different from
one, the pendulum orbits have four eigenvalue pairs different from one. We
considered four dof trace formula to describe simultaneously the quantization
of the axial and pendulum orbits. We considered four perpendicular dofs to
internuclear axis for pendulum orbits (two non-trivial different pairs of the
eigenvalues) and for axial orbits (four non-trivial pairs of the eigenvalues
where two pairs are distinct) in the trace formula. The two dof parallel to
internuclear axis are related to unstable motions (exception to $ZZee$ PO) are
not included in trace formula.

The related semiclassical density of states in four degrees of freedom for $R=2$
is presented in Figs.6. In the abscissa of Figs.6 one will find
$(-E_{el})^{-1/2}$ or the effective quantum number $(N_{eff})$ while the
ordinate shows the density of levels $[d(E_{el})]$ in arbitrary unity given by
a solid line. The vertical lines in these figures are the quantum eigenvalues
(discussed below)and the dotted line is a non-normalized smoothed quantum
spectrum. The quantum smoothed spectrum were resolved with the maximum orbital
period ($T_{pend}$).The smoothed quantum spectrum is given by

\begin{displaymath}
d_{Q}(E) = (\Delta E \sqrt{2 \pi})^{-1} \sum_{n}
e^{-(E-E_{n})^{2}/(2 \Delta E^{2})},
\end{displaymath}
\ \\
where $E_{n}$ are the quantum eigenvalues and $\Delta E = \hbar / T_{pend} =
1 / T_{pend}$. The smoothed quantum spectrum for all states is shown in Fig.6.

\begin{figure}
\begin{center}
\vskip 10pt
\includegraphics[width=15cm, height=15cm]{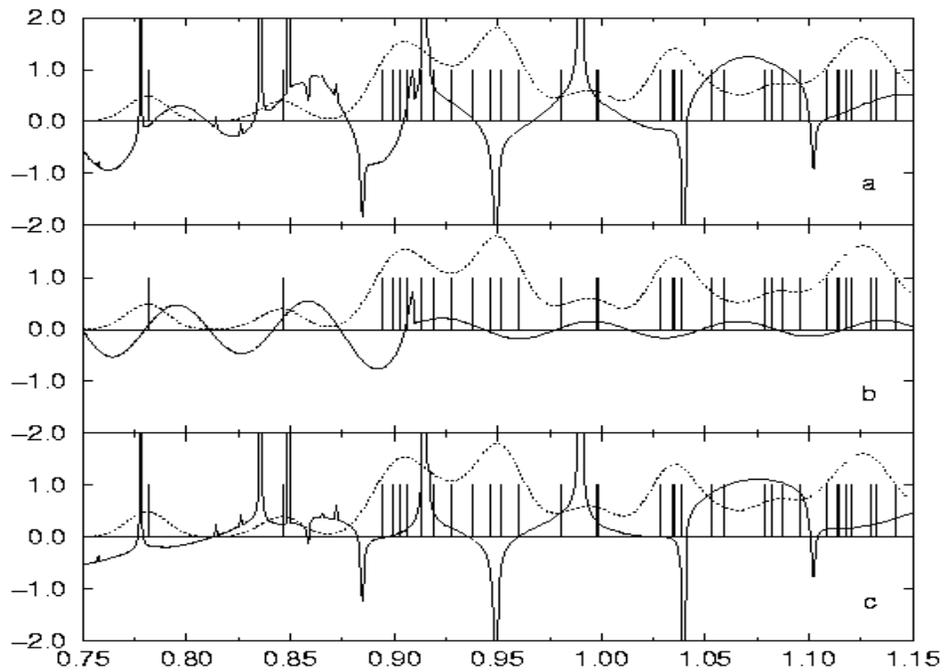}
\caption{Density of States in arbitrary unity as a function of effective quantum
number $Rez=(-E_{el})^{-1/2}$ (solid line) with R=2 a.u., quantum spectrum
(vertical lines) and smoothed quantum spectrum (dotted line). a) Density of
States for the axial and the pendulum orbits; b) for the pendulum orbits only;
c) for the axial ones only}
\end{center}
\end{figure}

The spectra calculated by Gutzwiller formula, presented in Fig.6, are smoothed
since they did not include many repetitions. The trace formula summed all
primitive orbits, but in this calculation only the Wannier (SS) and the
fundamental (AS) orbits (pendulum and $ZeeZ$ and $eZZe$ axial configurations)
and the molecular FPC were included. The semiclassical calculations take into
account a semi-quantitative agreement with quantum data for present degree of
approximation.

The Fig.6a shows the density of levels for simple sum of the pendulum and axial
orbits for $R=2$. The Fig.6b shows the density of levels for pendulum orbits
and the Fig.6c shows the density for axial orbits.

The study of the H$_2$ molecule is close to that for He atom. By analogy, we
consider the Wannier PO (SS) and the fundamental PO (AS) as first POs of the a
"symbolic dynamic", the SS PO is denoted by symbol $+$ and the AS PO is denoted
by symbol $-$. The symbolic dynamic is applied to the hyperbolic systems, but
the H$_2$ system is a mixed one and a symbolic dynamic can not describe the
whole system. Some parts of the H$_2$ can be described as a hyperbolic system
and other ones as near a integrable problem. All POs considered here show
generally an unstable behavior. We do not make any geometric series expansion
for trace formula and because of that, triplet states appear together with
singlet states in the density of levels.

It is necessary to consider a large number of POs in order to obtain a
semiclassical quantization with the trace formula and the zeta function.
However, a incipient semiclassical quantization can be given only few POs. The
trace formula and the zeta function were obtained for several configurations
simultaneously, i.e., the axial ($ZeeZ$, $eZZe$ and $ZZee$) and pendulum
configurations are convolutionated in the same trace formula and zeta function.
Since the whole H$_2$ system in 3-D connects all configurations that convolution
can be justified. For example, the Bohr and Langmuir's POs for H$_2$~\cite{ALC}
are connected to 1-D POs by eccentricity parameter. The Bohr and Langmuir's POs
belong to middle plane (or to plane that contain the internuclear axis) are
reduced to pendulum (or axial) POs for high value of the eccentricity.

The H$_2$ molecule has several important configurations where each one
contributed with infinity PO. Different from H$_2$, the He atom has only one
configuration with infinity PO. If one can describe the H$_2$ molecule it is
necessary to sum the PO for all configurations. There are mixes of those PO for
finite $R$. The H$_2$ system is reduced to He atom for $R \rightarrow 0$ and to
H$^+$ plus H$^-$~\cite{GR} or to H for $R \rightarrow \infty$. We make a simple
sum in trace formula, we do not consider any distinction among different
topologies of the 1-D configurations.

We calculated the Density of States for ground-state (singlet) and the first
triplet state with several geometries, i.e., with internuclear distance equal to
1.2, 1.6, 2.0 and 2.4au. These results are shown in Fig.7. The trace formula can
be calculated for any geometry of the H$_2$ system. These results are shown in
Fig.7 for ground and first triplet states with several internuclear distances.
These results agree very well with quantum data for ground state for $R=1.2$ and
$R=1.6$ and for other geometries and for first triplet state the agreement is
reasonable. The resonance $(N_{eff}=)0.836$ in Fig.7c is not important since it
appears due to the truncation process of the repetition and for slight change
this peak vanishes. Those agreement is relevant if we compare with the single
quantization results for ground and low excited states. The H$_2$ system can not
be described semi-quantitatively by a single quantization procedure since all
configurations contributed for those states.

\begin{figure}
\begin{center}
\vskip 10pt
\includegraphics[width=15cm, height=15cm]{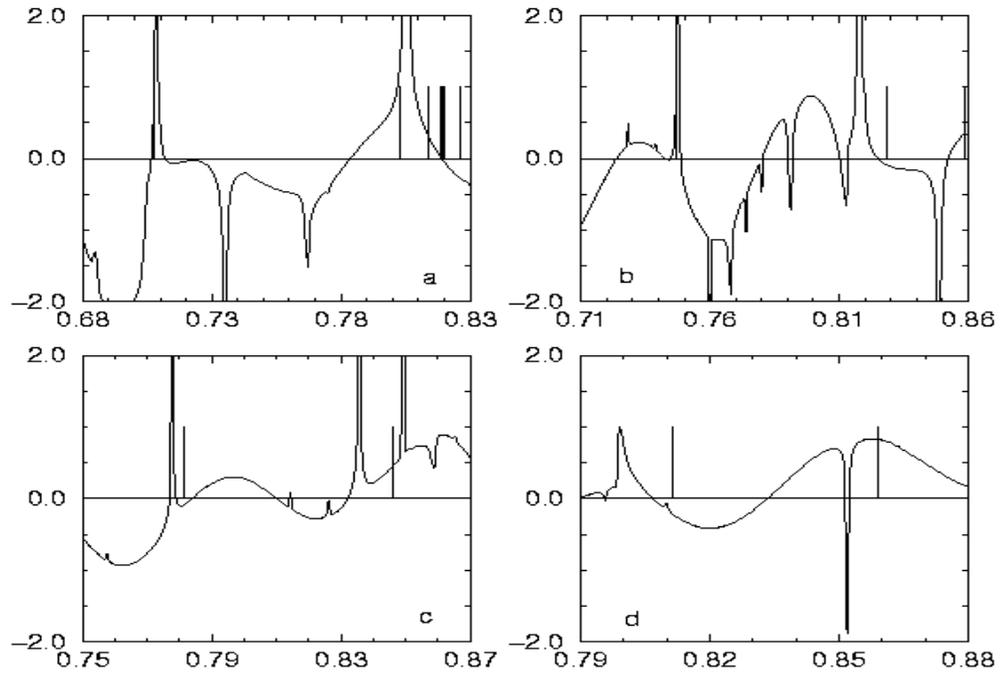}
\caption{Density of States in arbitrary unity for the axial and the pendulum
orbits as a function of $Rez$ (solid line) and quantum spectrum (vertical
lines). a) R=1.2au; b) R=1.6; c) R=2.0 and d) R=2.4}
\end{center}
\end{figure}

\begin{figure}
\begin{center}
\vskip 10pt
\includegraphics[width=15cm, height=15cm]{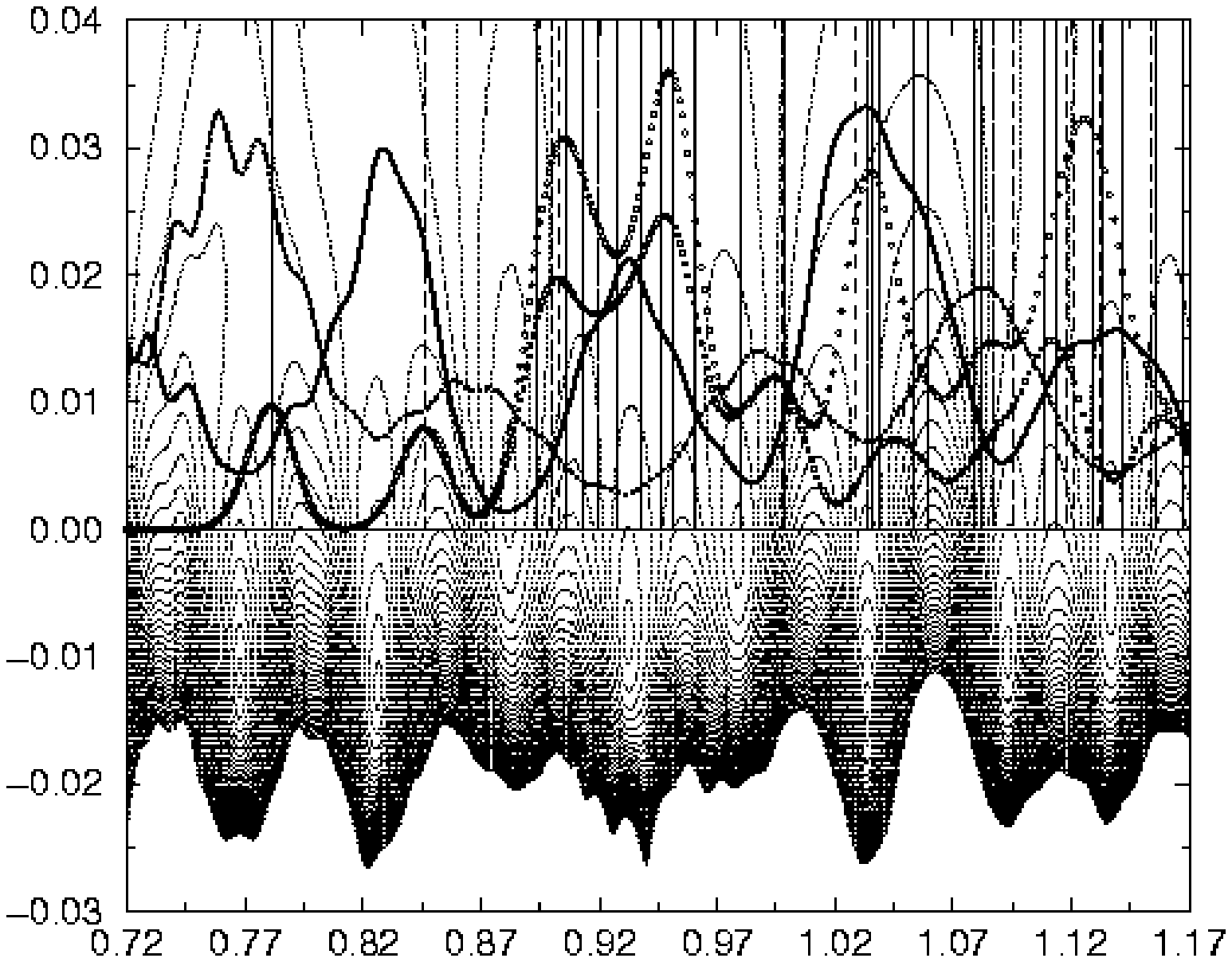}
\caption{Zeta function level curves for all POs as a function of the real
(abscissa) and imaginary (ordinate) part of the effective quantum number-$z$
(dotted lines) with R=2 a.u.; symmetric zeta function $(\zeta_{+}^{-1})$ square
module in arbitrary unity as a function of $Rez$ (solid line); antisymmetric
zeta function $(\zeta_{-}^{-1})$ (dotted line); quantum spectrum for (vertical
lines) singlet states (solid lines) and triplet states (dashed lines); smoothed
quantum spectrum for all states (circle) and without $\Pi$ and $\Delta$ states
(square)}
\end{center}
\end{figure}

\begin{figure}
\begin{center}
\vskip 10pt
\includegraphics[width=15cm, height=15cm]{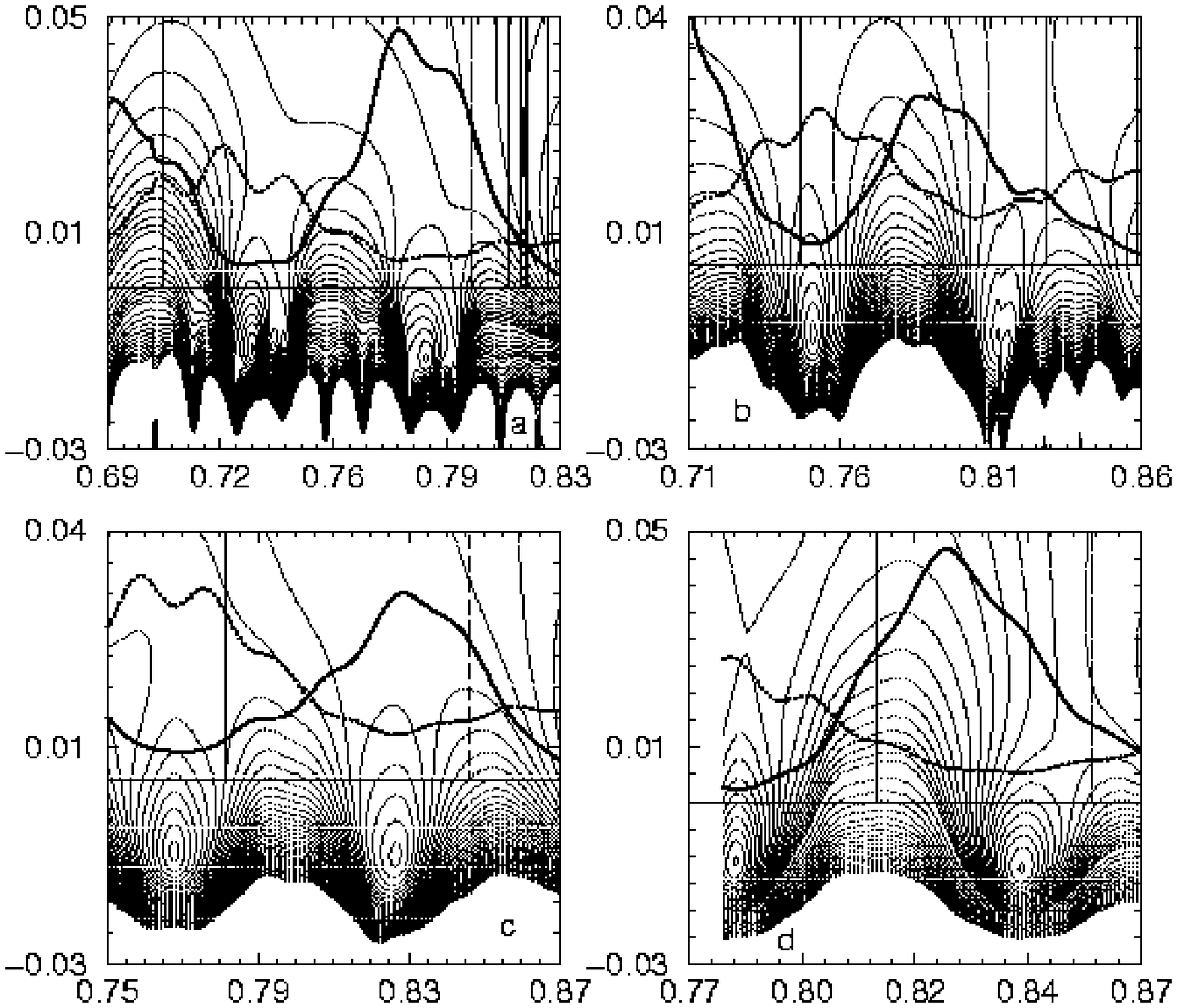}
\caption{Zeta function level curves for all POs as a function of the real
(abscissa) and imaginary (ordinate) part of the effective quantum number-$z$
(tinny dotted lines); symmetric zeta function $(\zeta_{+}^{-1})$ square module
in arbitrary unity as a function of $Rez$ (solid line); antisymmetric zeta
function $(\zeta_{-}^{-1})$ (dotted line) and quantum spectrum (vertical lines:
singlet states - solid lines, triplet states - dashed lines). a) R=1.2au; b)
R=1.6; c) R=2.0 and d) R=2.4}
\end{center}
\end{figure}

We can compare the configurations of the H$_{2}$ among each other and with the
configurations of the He atom. For example, the $eZZe$ configuration can
describe the $eZe$ one of the He for $R \rightarrow 0$. The $eZe$ configuration
is important to describe the atomic bounded states, but the ground state and
other excited states of the bounded H$_{2}$ system need more than that. For
example, the ground state of H$_{2}$ needs a bounded state between two hydrogen
atoms that it is not present in He system (the alpha particle is bounded!), the
H$_{2}$ is considered in adiabatic approximation. Perhaps, the $ZeeZ$ and
pendular configurations play the role of an essential task to describe the
ground-state.

That aspect becomes more complex the H$_{2}$ system than He and, in addition,
the lost of the spherical symmetry for H$_{2}$ system divides the importance
between the $ZeeZ$ and pendular configurations which does not exist in He atom.

The Fig.4 shows the correlation among the configurations for the united atom
limit and separated atoms limit with molecular configurations. All axial
configurations are also correlated to the $(Z+e+Ze)$ and $(Z+e+Z+e)$ separated
atoms limit. An important goal can be obtained the complete correlation between
the quantum states of the atomic limits with those of the molecular
configurations. For example, the separated atom limit of the ground-state and
first triplet state are integrate systems, i.e., two hydrogen atom in
ground-state described by wavefunctions $\Psi_1$ and $\Psi_2$. The simple sum
(linear combination of the atomic orbitals) of the wavefunction gives the
ground-state $(\Psi_{GS} = \Psi_1 + \Psi_2)$ and first triplet state
$(\Psi_{FTS} = \Psi_1 - \Psi_2)$. Those states coalesce to He states, i.e., for
respective ground-state and the first triplet state of the He. That na\"{\i}ve
description must be improved considering the correlation motion that is
implicit in the POs, which are absent in the simple sum of the wavefunction
shown above.

The agreement between the quantum spectrum and the density levels (Fig.6) is
weak for whole excited states since was considered few POs. It is necessary to
consider more POs with bigger period and also the stability islands. However, 
the ground-state and the first triplet state can be described better with these
few POs.

The ground-state for low values of the internuclear distance $(R=1.2)$ is
described mainly by $ZZee$ PO. That PO describes a repulsive behavior between
the nuclei (the nuclear repulsion dominates) that causes the expansion of the
molecule to equilibrium geometry $(R=1.4)$. The $ZeeZ$ POs dominate for increases
values of $R(=1.4 - 2.0)$, for bigger value of $R(=2.4)$ the pendulum orbits
increase their contributions. The electron attraction of the nuclei dominates for
$R>1.4$ maintaining the molecular binding.

The electrons are inside of the molecule for $R \geq 1.4$, i.e., with high
contribution of the $ZeeZ$ POs, which describe the bonded state. The first
triplet state is described mainly by $eZZe$ POs and few contribution of the
repetitions of the $ZZee$ PO for any geometry $(1.2 \leq R \leq 2.4)$. The
electrons of the $eZZe$ POs are outside of the molecule that characterizes a
non-bonded molecular state.

An important feature of the trace formula is: all contributions for ground state
resonance are obtained by repetitions of the POs, i.e., successive excitations of
the primitive PO. However, no important repetitions are made for $eZZe$ PO to
describe the first triplet state.

The zeta function gives the $eZZe$ configurations as the most important to
describe the (smoothed) quantum spectrum. The basic electronic structure can be
described by $eZZe$ configuration as $eZe$ configuration for He atom. However,
the $eZZe$ configuration fails to describe the bonding state and other
configurations must be added (see discussion above).

The zeta functions are shown in Fig.8. The zeta function level curves in complex
plane for all POs with R=2 a.u. is shown in Fig.8. That zeta function is
obtained from the sum of the zeta functions for each configuration. The
symmetric $(\zeta_{+}^{-1})$ and antisymmetric $(\zeta_{-}^{-1})$ zeta functions
square module in Fig.8 are given by product of the respective zeta functions for
each configuration. The quantum spectrum (vertical lines: singlet states - solid
lines, triplet states - dashed lines), the smoothed quantum spectrum for all
states (circle) and without $\Pi$ and $\Delta$ states (square) are also disposal
in Fig.8. The $\zeta_{+}^{-1}$ gives the singlet resonances and $\zeta_{-}^{-1}$
gives the triplet ones. All resonances (minimum) obtained by zeta functions are
correlated with peaks of the smoothed quantum spectrum.

The results for trace formula and zeta function for $R=2.0$ are summarized in
Table.I.

TABLE I. Effective quantum number for quantum (eigenvalues) and semiclassical
(resonances) calculations with $R=2.0$: $SQS$ - smoothed quantum spectrum;
$SQS^*$ - smoothed quantum spectrum without $\Pi$ and $\Delta$ states; $d(E)$ -
trace formula; $\zeta^{-1}$ - resonances for $\zeta_{+}^{-1}$ and
$\zeta_{-}^{-1}$; $\zeta^{-1 *} = \zeta_{ZeeZ}^{-1} + \zeta_{eZZe}^{-1} +
\zeta_{ZZee}^{-1} + \zeta_{pendulum}^{-1}$

\begin{eqnarray*}
\begin{array}{ccccc}
SQS & SQS^* & d(E) & \zeta^{-1} & \zeta^{-1 *} \\
\\
0.781 & 0.781 & 0.777 & 0.767 & 0.768 \\
0.846 & 0.846 & 0.849 & 0.824 & 0.827 \\
0.904 & 0.902 & --    & 0.879 & 0.882 \\
0.949 & 0.948 & --    & 0.932 & 0.932 \\
0.994 & 0.994 & --    & 0.986 & 0.979 \\
1.035 & 1.045 & --    & 1.035 & 1.034 \\
1.085 & 1.112 & --    & 1.090 & 1.094 \\
1.125 & 1.159 & --    & 1.137 & 1.137 \\
\end{array}
\end{eqnarray*}

\ \\

The zeta functions were calculated for several geometries and these results are
shown in Fig.9. The resonances of the trace formula and the zeta function for
ground state are summarized in Table.II and for first triplet state in
Table.III.

TABLE II. Effective quantum number $(N_{eff})$ for ground state resonances for
several geometries. The numbers in bracket are the total energy
$(E_t=1/R-1/N_{eff}^2)$

\begin{eqnarray*}
\begin{array}{cccc}
R & Quantum & d(E) & \zeta^{-1} \\
\\
1.2 & 0.7074 (-1.1650) & 0.708(2) (-1.1605) & 0.731 (-1.038) \\
1.4 & 0.7277 (-1.1741) & -------            & 0.742 (-1.102) \\
1.6 & 0.7467 (-1.1685) & 0.747(0) (-1.1671) & 0.752 (-1.143) \\
2.0 & 0.7813 (-1.1382) & 0.777(3) (-1.1551) & 0.769 (-1.191) \\
2.4 & 0.8114 (-1.1022) & 0.798(8) (-1.1505) & 0.780 (-1.227) \\
\end{array}
\end{eqnarray*}

\ \\

TABLE III. Effective quantum number $(N_{eff})$ for first triplet state
resonances for several geometries. The numbers in bracket are the total energy
$(E_t=1/R-1/N_{eff}^2)$

\begin{eqnarray*}
\begin{array}{cccc}
R & Quantum & d(E) & \zeta^{-1} \\
\\
1.2 & 0.8026 (-0.7191) & 0.805(1) (-0.7094) & 0.779 (-0.815) \\
1.4 & 0.8169 (-0.7842) & -------            & 0.794 (-0.872) \\
1.6 & 0.8285 (-0.8319) & 0.818(1) (-0.8691) & 0.804 (-0.922) \\
2.0 & 0.8460 (-0.8972) & 0.849(1) (-0.8870) & 0.827 (-0.962) \\
2.4 & 0.8592 (-0.9379) & 0.857(3) (-0.9439) & 0.843 (-0.990) \\
\end{array}
\end{eqnarray*}

\ \\

The ground state shows a minimum in total energy for $R=1.4$. The trace formula
gives same feature, in despite of the calculation problem for $R=1.4$ (see
discussion above). The zeta function does not give the minimum and the state has
a repulsive behavior as WKB calculations. The repulsive behavior of the first
triplet state is obtained for all semiclassical calculations. The calculation
with trace formula for $R=1.4$ gives a different result. The effective quantum
number for resonances of the ground state is $0.724 (E_t=-1.195)$ and for first
triplet state is $0.819 (E_t=-0.775)$ in absence of the $ZZee$ and pendular POs.
If those POs are included on the trace formula the resonances are destroyed for
the two states.

How the $eZZe$ configuration dominates the results of the zeta function then the
resonances are near to those of the WKB results for that $eZZe$ POs. The 1-D
configurations are not linked to each other, only in 2-D and 3-D, and the
convolution is critical to maintain the balance among the configurations. For
example, if we change the contribution of the $eZZe$ configuration for half, the
resonances of the ground and first triplet states shift to similar results of
the trace formula. The pendulum, $ZZee$ and $ZeeZ$ configurations contribute to
describe better the values of the semiclassical resonances for ground-state,
giving the minimum (bounded state), and first triplet state. The trace formula
indicates that the ground state is described mainly by successive excitations of
some POs. It indicates that zeta function can need POs with large period to
reproduce the resonances with same accuracy of the trace formula for ground
state and first triplet one.

The semiclassical quantization of the H$_2$ molecule is remarkable complex.
However, some important features can be obtained. The results of the trace
formula and zeta function are complementary for few POs. The trace formula gives
some sharp results, but appears frequently spurious resonances and complicated 
profile. The zeta function is better behaviored than trace formula, but some
difficulties appear, e.g., when several configurations are considered. A better
result is managed if two semiclassical methods are considered
complementarity.

The semiclassical quantization considered in these studies was reduced to the
perpendicular plane to the internuclear axis. Some considerations allow that
reduction, the parallel motion to chemical bond is unstable and the
perpendicular is stable or weakly unstable depending on the energy. That
reduction can be better understood when the chemical bonding will be discussed.
The main plane to describe the chemical bonding is the perpendicular one, but
for non-bonding state perhaps the parallel plane can be important.

\subsection*{Quantum studies}

The Schr\"odinger equation for H$_{2}$ molecule was calculated using the MELD
program~\cite{MELD}. This program uses as initial guess a linear combination  of
the Cartesian Gaussian atomic bases. These atomic bases were constituted by
$9s$, $6p$, $4d$, $3f$, and $2g$ atomic function, where $s,p,d,f,$ and $g$
represent the usual notation of the hydrogen atom solution. The Cartesian
Gaussian has the angular part described by a polynomial, e.g., for $s$ function
the polynomial is degree zero, for $p$ is degree one, etc. The electronic
structure was described at the full configuration-interaction level and having
the Hartree-Fock configuration as the zero-order function. The lowest
eigenvalues and eigenvectors were obtained of the configuration-interaction
matrix for all symmetries for the $D_{2 h}$ point group  and for singlet and
triplet multiplicity. The calculations were performed in $D_{2 h}$ subgroup of
the $D_{\infty h}$, because of the limitation of the program.

The quantum eigenvalues are given in Table.IV in order to compare the accuracy
of the quantum and semiclassical calculations. The complete spectra of the
singlet and triplet states are shown in Fig.10.

\ \\

TABLE IV. Electronic energy for several states of the H$_{2}$ for different
symmetries and geometries.

\begin{eqnarray*}
\begin{array}{cccccc}
States           & R=1.2       & R=1.4       & R=1.6       & R=2.0       &
R=2.4 \\
\\
^1\Sigma_{g}^{+} & -1.99807482 & -1.88858658 & -1.79342268 & -1.63799147 &
-1.51896199 \\
^3\Sigma_{u}^{-} & -1.55233372 & -1.49839961 & -1.45669112 & -1.39704906 &
-1.35456212 \\
^1\Sigma_{u}^{-} & -1.49369920 & -1.41915299 & -1.35496600 & -1.25143196 &
-1.17284326 \\
^3\Sigma_{g}^{+} & -1.50929781 & -1.42723452 & -1.35522921 & -1.23551307 &
-1.14076781 \\
^3\Pi_{u}        & -1.48957702 & -1.41139779 & -1.34262235 & -1.22772023 &
-1.13605465 \\
^1\Sigma_{g}^{+} & -1.48676404 & -1.40583131 & -1.33492395 & -1.21739177 &
-1.12506771 \\
^1\Pi_{g}        & -1.46419102 & -1.38464740 & -1.31480857 & -1.19862781 &
-1.10669065 \\
^3\Sigma_{u}^{-} & -1.43183076 & -1.35736416 & -1.29197876 & -1.18282088 &
-1.09598014 \\
^1\Sigma_{u}^{-} & -1.41374173 & -1.33760497 & -1.27141851 & -1.16262421 &
-1.07773494 \\
^3\Sigma_{g}^{+} & -1.40640301 & -1.32554820 & -1.25448492 & -1.13605979 &
-1.04249061 \\
^1\Sigma_{g}^{+} & -1.38661492 & -1.30571704 & -1.23474544 & -1.11695144 &
-1.02912903 \\
\end{array}
\end{eqnarray*}

\ \\

The quantum calculation performed considers that all states are discrete, which
represent an important limitation to the calculation. Several quantum methods
can  be applied to calculate the H$_{2}$ system, see subsection Quantum Methods.

\begin{figure}
\begin{center}
\vskip 10pt
\includegraphics[width=15cm, height=17cm]{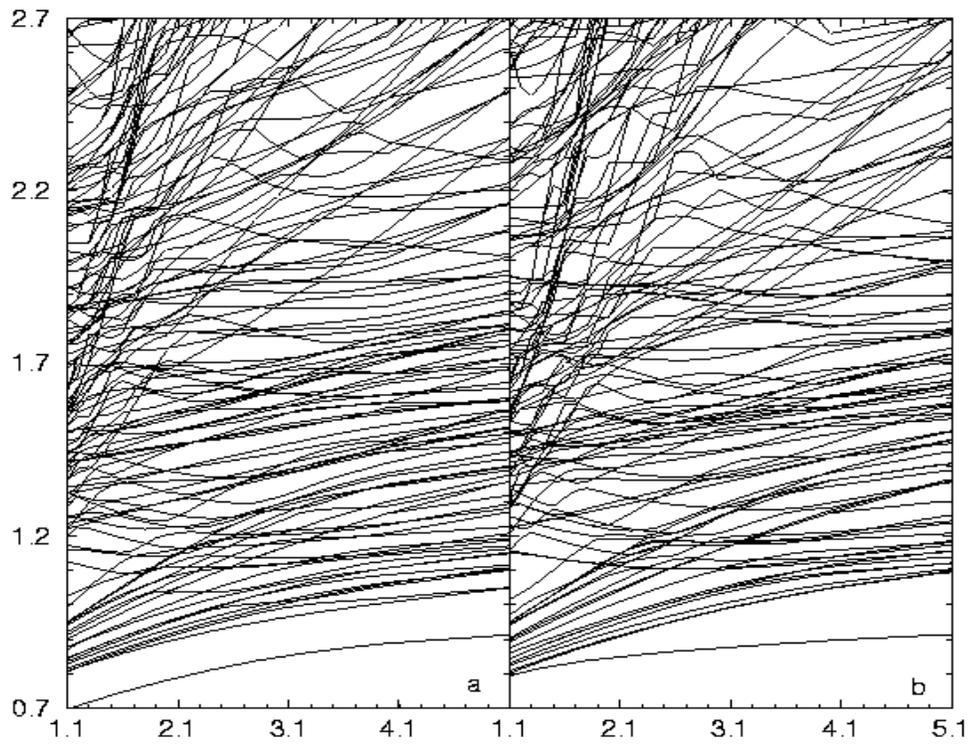}
\caption{Effective quantum number $(N_{eff}=(-E_{el})^{-1/2})$ as a function
of internuclear distance $(R)$. a) singlet states; b) triplet states} 
\end{center} 
\end{figure}

The difficulties of the semiclassical calculation of the H$_2$ molecule can be
better understood with Fig.10. The numerous avoid-crossings are present for
excited states. Those avoid-crossings interchange the electronic configurations
among the states. If we disposal a significant numbers of the POs it will be
possible to describe those complex spectra.

Correlation between the classical orbit and the shape of the probability density
of the molecular orbitals can also be managed~\cite{ALC1}. The appearance of 
scars in the wavefunction can help to find specific POs. Of course, a systematic
search of the POs is necessary for complete knowledge of the H$_2$ system.

We can also use some chemical knowledge to find some classical track in the
multidimensional quantum world. Those knowledge can be the electronic
distribution around the chemical bonding; electroafinity, i.e., localization of
the electron around some preferential atoms; adiabatization coordinates
(repulsion and attraction coordinates), chemical bonding dynamics (rate between
kinetic and potential energies), etc. One can select some periodic orbits
considering the molecular orbital shape of the states, e.g., $\sigma$ states
with Bohr orbits and $\pi$ states with Langmuir orbits~\cite{ALC1}; covalent
states with $ZeeZ$ and pendulum orbits; repulsive states with $eZZe$; ionic
states with $ZZee$ orbits; the profile of $scars$; etc.

\subsection*{The Chemical Bonding}

Ruedenberg~\cite{KR} showed that the chemical bond is formed by decreasing of
kinetic energy in the chemical bond direction. That aspect can be better
understood with classical and semiclassical physics. It is possible to infer
that the perpendicular plane to the axial direction is the most important
region to describe the chemical bonding since the depression of the kinetic
energy provokes the accumulation of electrons around the perpendicular plane
of chemical bond axis.

H$_{2}$ has basic ingredients to study the chemical bond, i.e., a pair of
electrons. Perhaps, there are connections between the Rudenberg's theory and the
classical mechanics of the POs.

It is possible to verify the correlation between the stability of the PO of the
H$_{2}$ and the chemical bonding formation. It is known the origin of the
bonding  energy of the H$_{2}$ molecule is a consequence of the kinetic energy
depression that happens with accumulation of the electronic charge in the
internuclear region~\cite{KR,HSBC}. It is possible to anticipate some
manifestations of chemical bonding properties with 1-D orbits. The instability
of the 1-D PO in the internuclear direction indicates that the motion of the
electrons in that direction can not be important in the formation of the
chemical bonding. Perpendicular motion to internuclear axis, that is a stable
motion or weakly unstable one, can be important to formation of the chemical
bonding since the electrons can orbit the nuclei in a plane perpendicular to
internuclear axis. Therefore, Bohr's, Lewis', and other equivalent theories of
the electron pair localization~\cite{HK} can have a simple interpretation  with
a deep meaning in the description of the chemical bonding.

The H$_{2}$ molecule has some return points near to nuclei (centrifugal barrier)
and near to average middle point of the internuclear axis (electron-electron
repulsion). The average kinetic energy near to nuclei is high and near to middle
point is low. The electronic charge can accumulate dynamically near to middle
point because of the slow motion and consequently the electron permanence is
high in average. The permanence time depends on the motion, on the PO and on the
state (bonding or not).

We studied the behavior of the kinetic energy of the $ZeeZ$ and pendulum
fundamental POs and verify that there are several regions of the electron
accumulation.

A simple connection between the Bohr orbit of the H$_2$ and Ruedenberg's theory
of the origin of the binding energy for H$_2$ was found~\cite{HSBC}. That
connection considers the behavior of the kinetic energy and the virial theorem
for some variations over the Bohr orbit.

Shortly, Ruedenberg's work~\cite{KR,WK} shows that the chemical bond formation
is consequence of the depression of the kinetic energy along of the internuclear
axis. The module of the total $|E_t|$, potential $|E_p|$ and kinetic $|E_k|$
energies of the molecule are bigger than those of the separated atoms with
$|E_t|=|E_k|=|E_p|/2$ (virial rate). The molecule can be seen as a box with
dimension along of the internuclear axis $(\Delta X_2)$ bigger than the atom
$(\Delta X_1)$ and the dimension perpendicular to the internuclear axis $(\Delta
X_2^p)$ smaller than the atom $(\Delta X_1^p)$. That is, $\Delta X_1 < \Delta
X_2 \Rightarrow \Delta P_1 > \Delta P_2$ and $\Delta X_1^p > \Delta X_2^p
\Rightarrow \Delta P_1^p < \Delta P_2^p$. How $\Delta P_1 > \Delta P_2$, it is
more probable to obtain a smaller kinetic energy in the parallel direction of
the chemical bond of the molecule than of the atom; conversely for perpendicular
direction. It is possible to show that the smaller value of the kinetic energy
causes the depression in it around the middle plane and the electrons can
accumulate between the nuclei. The electron accumulation and the stable motion
in the perpendicular direction of the internuclear axis are the classical
ingredients of the chemical bond.

Now, we need to obtain the principal POs that describe the ground state in order
to study the formation of the chemical bond.

Our semiclassical studies give us that the $eZZe$ POs contribute to describe as
a quasi-classical interaction (each electron is described by an independent
electronic density)~\cite{WK} of the two hydrogen atom to constitute the ground
state of the molecule. That $eZZe$ configuration is enough to describe the
united atom limit, i.e., He atom, but it is not for the chemical bond since the
$eZZe$ POs give a non-bonding behavior.

The chemical bond is obtained by quantum interference of the atomic wavefunction
$(\Psi_{GS} = \Psi_1 + \Psi_2)$. That interference can be mimicked by the
semiclassical quantization since it describes the minimum of the energy for
ground state. Bohr quantization for H$_2$ gave the first calculation of the
chemical bond~\cite{NB,HK}. The connection of that (semiclassical single)
quantization with Ruedenberg's theory was made~\cite{HSBC}. However, Bohr
quantization gave a qualitative result since the H$_2$ molecule can not be
reduced to one simple PO. The H$_2$ molecule is a complex system, several POs
must participate of a (semi)quantitative description.

The ground state of the H$_2$ is described basically by $eZZe$, $ZeeZ$ and
pendulum POs. The $eZZe$ POs are very important since gives the principal
interaction of the charges. The $ZeeZ$ and pendulum POs give the fine adjust in
order to manage the chemical bond.

The $ZeeZ$ and pendulum POs were studied classically and they showed similar
depression of the kinetic energy obtained with quantum mechanics~\cite{KR}.
Probably the POs which show a depression of the kinetic energy between the
nuclei are important to describe a chemical bond. The minima of the potential
energy for excited states caused by avoid-crossings (see Fig.10) can be
different.

The starting point of the studies of the semiclassical quantization of the
H$_2$~\cite{ALC,ALC1} was the Bohr orbit for H$_2$~\cite{NB}. Bohr extended that
model (orbit) to describe other molecules. That Bohr's idea was applied to
several molecules~\cite{RDH} and potentially the semiclassical quantization can
also be applied to those systems.

In summary, some 1-D POs show an accumulation of the electronic charge  between
the nuclei. The stability of the perpendicular dof of the internuclear axis
shows a stable or weakly unstable motion. Therefore, the electrons can
accumulate preferencially in the chemical bond field.

\section{Conclusions}

Several studies for the H$_{2}$ molecule were shown in this paper. These studies
are important for a future understanding of the connection between classical and
quantum worlds (``the Correspondence Principle'') to chaotic problems and of the
chemical bond since the H$_{2}$ system contains the basic ingredients
(two-electron binding) for that.

Some Bohr-Sommerfeld quantizations have been performed, but those calculations
were not enough to describe the $H_{2}$ system. Unfortunately, that (chaotic)
system is not integrable and a global semiclassical quantization must be made.
Gutzwiller formula and the zeta function were calculated for 1-D periodic
orbits.

We can enumerate some perspectives of this work:

a) Obtain the relation between chemical bond and classical properties of
periodic orbits: the POs which show a depression of the kinetic energy between
the nuclei are important to describe a chemical bond;

b) Verify the Correspondence Principle for H$_{2}$, mainly in chaotic regime;

c) Use the classical and semiclassical mechanics to help the quantum mechanic
interpretation;

d) Use the semiclassical method in order to give the correct results for high
excited state in substitution of the quantum numerical calculation.

Mainly, this work describes a theoretical study of the non-linear dynamics of
the hydrogen molecule. We show a first step to understand ``the Correspondence
Principle'' in the chemical world that treats essentially nonlinear systems.

\section*{Acknowledgments}

Acknowledgments are given for Dr.\ Gregor Tanner, Prof.\ C\'esar de Oliveira,
Dr.\ Francisco Correto Machado, Prof.\ Rog\'erio Cust\'odio and Prof.\ Carles
Sim\'o for fruitful discussions and for Dr.\ Nenad Simonovi\`c to provide his
program discussed in reference~\cite{NSS}. The author acknowledges specially
Dr.\ Gregor Tanner, University of Nottingham and Hawlett-Packard Labs (Bristol)
to provide support to visit to England. Financial support by FAPESP (Brazil) is
acknowledged.

\end{document}